\documentclass[times, tight, twocolappendix]{aastex631} % 
\usepackage{amsfonts}
\usepackage{mathrsfs}
\usepackage{lineno}
\usepackage{multirow}
\usepackage{hyperref}
\hypersetup{
    colorlinks=true,
    linkcolor=blue,
    citecolor=blue,
    urlcolor=blue
}

\newcommand\cm{\checkmark}
\newcommand{\cs}{c_{\rm s}}
\newcommand{\dotMp}{\dot{\mathscr{M}}_{\rm p}}
\newcommand{\dotMs}{\dot{\mathscr{M}}_{\rm s}}
\newcommand{\etaout}{\eta_{\rm out}}
\newcommand{\etatwo}{\eta_{\bar{3}}}
\newcommand{\etathree}{\eta_{\bar{3}}}
\newcommand{\etafour}{\eta_{\bar{4}}}
\newcommand{\facc}{f_{\rm acc}}
\newcommand{\gammax}{\gamma_{\rm max}}
\newcommand{\gammin}{\gamma_{\rm min}}
\newcommand{\Hp}{H_{\rm p}}
\newcommand{\Hs}{H_{\rm s}}
\newcommand{\kB}{k_{\rm B}}
\newcommand{\Lout}{L_{\rm out}}
\newcommand{\Lpeak}{L_{\rm peak}}
\newcommand{\mathdotM}{\dot{\mathscr{M}}}
\newcommand{\me}{m_{\rm e}}
\newcommand{\MBon}{M_{\rm Bon}}
\newcommand{\Mcav}{M_{\rm cav}}
\newcommand{\Mout}{M_{\rm out}}
\newcommand{\Mp}{M_{\rm p}}
\newcommand{\Ms}{M_{\rm s}}
\newcommand{\Msun}{M_\odot}
\newcommand{\Ps}{P_{\rm s}}
\newcommand{\qq}{q_{\bar{6}}}
\newcommand{\qqfive}{q_{\bar{5}}}
\newcommand{\qqsix}{q_{\bar{5}}}
\newcommand{\RBon}{R_{\rm Bon}}
\newcommand{\Rcav}{R_{\rm cav}}
\newcommand{\Rg}{R_{\rm g}}
\newcommand{\rg}{r_{\rm g}}
\newcommand{\RHill}{R_{\rm Hill}}
\newcommand{\Rout}{R_{\rm out}}
\newcommand{\rhop}{\rho_{\rm p}}
\newcommand{\rhos}{\rho_{\rm s}}
\newcommand{\Sigp}{\Sigma_{\rm p}}
\newcommand{\sigSB}{\sigma_{\rm SB}}
\newcommand{\sigT}{\sigma_{\rm T}}
\newcommand{\Tc}{T_{\rm c}}
\newcommand{\Ts}{T_{\rm s}}
\newcommand{\Tsh}{T_{\rm sh}}
\newcommand{\Teff}{T_{\rm eff}}

\newcommand{\tcoolff}{t_{\rm cool}}

\newcommand{\tdiff}{t_{\rm diff}}

\newcommand{\teff}{t_{\rm eff}}
\newcommand{\tgrow}{t_{\rm grow}}
\newcommand{\tlast}{t_{\rm last}}
\newcommand{\tlind}{t_{\rm Lind}}
\newcommand{\tmig}{t_{\rm mig}}
\newcommand{\tout}{t_{\rm out}}
\newcommand{\trej}{t_{\rm rej}}
\newcommand{\tviss}{t_{\rm vis}}
\newcommand{\tvisp}{t_{\rm vis}^\prime}
\newcommand{\OmgK}{\Omega_{\rm K}}
\newcommand{\OmgKs}{\Omega_{\rm K,s}}
\newcommand{\vout}{v_{\rm out}}
\newcommand{\vrp}{v_{\rm r}^\prime}
\newcommand{\vrs}{v_{\rm r}}
\newcommand{\vturb}{v_{\rm turb}}
\newcommand{\Xout}{X_{\rm out}}
\newcommand{\yr}{\rm yr}
\newcommand{\gamlumOut}{10^{42}}
\newcommand{\gamlumMid}{10^{43}\rm \,erg\,s^{-1}}
\newcommand{\therlumOut}{10^{43}\rm \,erg\,s^{-1}}
\newcommand{\therlumMid}{10^{44}\rm \,erg\,s^{-1}}
\newcommand{\therlumInn}{10^{43}\rm \,erg\,s^{-1}}

\newcommand{\synlumInn}{10^{42}\rm \,erg\,s^{-1}}
\newcommand{\synlumMid}{10^{43}}
\newcommand{\Ep}{E_{\rm p}}
\newcommand{\Eg}{E_{\rm g}}
\newcommand{\Eout}{E_{\rm out}}

\def\ihep{Key Laboratory for Particle Astrophysics, Institute of High Energy Physics,
Chinese Academy of Sciences, 19B Yuquan Road, Beijing 100049, China}
\def\AstroUCAS{School of Astronomy and Space Sciences, University of Chinese Academy of Sciences, 
19A Yuquan Road, Beijing 100049, China}

\def\PhyUCAS{School of Physics, University of Chinese Academy of Sciences, 
19A Yuquan Road, Beijing 100049, China}

\def\naoc{National Astronomical Observatory of China, 20A Datun Road, Beijing 100020, China}

\shorttitle{Accretion-modified Stars in AGNs}
\shortauthors{Liu et al.}

\begin{document}
\title{\large\bf
Accretion-modified Stars in Accretion Disks of Active Galactic Nuclei: Observational Characteristics in Different Regions of the Disks}

\author[0000-0003-3086-7804]{Jun-Rong Liu}
\affil{\ihep}
\affil{\PhyUCAS}

\author{Yi-Lin Wang}
\affil{\ihep}
\affil{\PhyUCAS}

\author[0000-0001-9449-9268]{Jian-Min Wang}
\affil{\ihep}
\affil{\AstroUCAS}
\affil{\naoc}

\correspondingauthor{Jian-Min Wang}
\email{wangjm@ihep.ac.cn}
% \linenumbers

\begin{abstract}
Stars and compact objects embedded in accretion disks of active galactic nuclei (AGNs), dubbed ``{\it accretion-modified stars}" (AMSs), often experience hyper-Eddington accretion in the dense gas environment, resulting in powerful outflows as the Bondi explosion and formation of cavities.
The varying gas properties across different regions of the AGN disk can give rise to diverse and intriguing phenomena.
In this paper, we conduct a study on the characteristics of AMSs situated in the outer, middle, and inner regions of the AGN disk,
where the growth of the AMSs during the shift inward is considered.
We calculate their multiwavelength spectral energy distributions (SEDs) and thermal light curves.
Our results reveal that the thermal luminosity of the Bondi explosion occurring in the middle region leads to UV flares with a luminosity of $\sim 10^{44}\rm \,erg\,s^{-1}$.
The synchrotron radiation of Bondi explosion in the middle and inner regions peaks at the X-ray band with luminosities of $\sim 10^{43}$ and $\sim 10^{42}\rm \,erg\,s^{-1}$, respectively.
The $\gamma$-ray luminosity of IC radiation spans from $10^{42}-10^{43}\rm \,erg\,s^{-1}$ peaked at the $\sim 10\,$MeV (outer region) and $\sim$ GeV (middle and inner regions) bands.
The observable flares of AMS in the middle region exhibit a slow rise and rapid Gaussian decay with a duration of months,
while in the inner region, it exhibits a fast rise and slow Gaussian decay with a duration of several hours.
These various SED and light-curve features provide valuable insights into the various astronomical transient timescales associated with AGNs.

\end{abstract}

\subjectheadings{Active galactic nuclei (16); Supermassive black holes (1663)}

\section{Introduction}
In light of high metallicity in broad-line regions (BLRs) of active galactic nuclei \citep[AGNs;][]{Hamann1999}, stellar evolution and compact objects in AGN accretion disks were originally realized by \cite{Artymowicz1993} and \cite{Cheng1999} for metal production in BLRs, $\gamma$-ray bursts (GRBs), and gravitational waves (GWs), respectively. 
There is fast-growing evidence for such a scenario of compact objects in the accretion disks from LIGO detections of GWs.
The detection of GWs from the merger of binary black holes (BBHs) with masses heavier than typical stellar BHs, particularly the GW190521 event \citep{Abbott2020},
has potentially indicated that AGN disks could serve as sites for BBH mergers. 
Utilizing observations from the Zwicky Transient Facility,
\cite{Graham2020} reported the first potential optical electromagnetic counterpart (EMC) candidate (ZTF19abanrhr) for the GW event S190521g,
which could originate from a BBH merger within an AGN disk.
Although the potential association between these two astronomical transients remains a subject of ongoing debate \citep[e.g.,][]{Ashton2021, Palmese2021, Morton2023}, there is emergence of more appealing evidence from 
\cite{Graham2023} who subsequently reported nine EMC candidates for BBH mergers detected by LIGO/Virgo during the O3 run.
Additionally, \cite{Lazzati2023}, \cite{Levan2023}, and \cite{Tagawa2023BBH} claimed several observed GRBs possibly occurring in an AGN disk.
\cite{LiFan2023} and \cite{Han2024} proposed that some GW events may originate from a BBH merger in AGN disks.
Moreover, it would be a giant step to understand AGN phenomena of fueling the central supermassive black holes (SMBHs) and metallicity \citep{Wang2010,Wang2011,Wang2012,Wang2023,Fan2023} if some LIGO GW events with association with AGNs are identified eventually.

Much theoretical attention has been paid to stellar evolution, compact objects in AGN disks, and related astrophysics in recent years, as summarized in Table \ref{tab:ref}.
There is widespread research on the evolution of single stars \citep{Cheng1999, Cantiello2021, Dittmann2021, McKernan2022, Jermyn2022, Ali-Dib2023} and the evolution of stellar populations via accretion in the dense environment \citep{Wang2023}.
Stars undergo supernova explosions \citep{Grishin2021, Li2023} in the final stage of their evolution or may even be tidally disrupted by the stellar-mass black hole \citep[sMBH;][]{Yang2022}, leading to metal enrichment \citep{Wang2023, Huang2023} of the AGN disk and the formation of compact objects.
The accretion onto compact objects influences their spin evolution \citep{Jermyn2021, Chen2023} and growth \citep{McKernan2012, Davies2020}, giving rise to abundant observational signatures, such as Bondi explosion from powerful outflows \citep{PaperI}, jet flare \citep{PaperII, Tagawa2023BH}, and the so-called ``{\it accretion-induced collapse}'' of neutron star \citep[NS;][]{Perna2021AIC} or white dwarf \citep[WD;][]{Zhu2021WD}.
Numerous compact objects in the AGN disk inevitably form binary systems \citep{Tagawa2020, Rowan2023, Wang2024} via Jacobi capture \citep{PaperII}, GW emission \citep{Rom2024}, dynamical friction \citep{Qian2024}, or the collision of their circum-single disks \citep{Li2023Jiaru}.
Simulations of BBH evolution show that the complex interactions between the BBH and AGN disk induce the contraction of the BBH separation \citep{Li2022, LiRixin2022, Dittmann2024}, leading to BBH mergers \citep{Bartos2017, Yang2019, Yi2019, McKernan2020, Samsing2022}, often followed by electromagnetic flares \citep{PaperII, Chen2024, Tagawa2023Shock}, such as GRBs \citep{Perna2021EMC, Lazzati2022, Yuan2022, Ray2023} and GRB afterglow \citep{Wang2022}.
Additionally,  WD collisions \citep{Luo2023, Zhang2023} and kilonova emission from NS-NS/BH-NS mergers \citep{Zhu2021NSmerger, Ren2022, Kathirgamaraju2023} are also important probes for studying compact objects in AGN disks.

As the black holes (BHs) in AGN disks are generally undergoing hyper-Eddington accretion ($\sim 10^9\,\dot{M}_{\rm Edd}$) and AMSs \citep[][]{PaperI},
where {\it stars} can widely refer to stars, WDs, NSs, or BHs.
In such a scenario, strong outflows develop \citep[e.g.,][]{Kitaki2021, PaperI}.
As a consequence, outbursts from the BHs emerge as the Bondi explosion through the interaction between powerful outflows and dense AGN disk gas \citep{PaperI, Tagawa2022, ChenKen2023a}, resulting in thermal and nonthermal radiations \citep[e.g.,][]{PaperI, Tagawa2023BH}.
\cite{PaperIII} first applied the AMS model to Sgr A$^*$ with low accretion rates, naturally reproducing the orbit observed by GRAVITY/VLTI \citep{GRAVITY2020, Gravity2023} and successfully explaining the period of the flare/flicking in the light curve \citep{Genzel2003, Genzel2010} and spectral energy distribution (SED) from radio to X-ray bands \citep{Witzel2021, Boyce2022}.
It is important to note that the above AMS is situated within advection-dominated accretion flows \citep[ADAFs;][]{Narayan1995ADAF, Narayan1995}.
Considering the environment of accretion BHs, we explore the AMSs located in the outer, middle, and inner regions of the standard disk \citep{Shakura1973},
leading to distinct physical processes and observational characteristics.

This is the fourth paper in a series aimed at investigating the behavior of BHs within the AGN disks \citep{PaperI, PaperII, PaperIII}.
This paper is structured as follows.
In \S\,\ref{sect:model}, we present our AMS model, including the modified Bondi accretion and the formation of cavities through the interaction of strong outflows with AGN disks.
\S\,\ref{sect:application} provides a straightforward application of the AMS model within the context of the standard disk (inner and middle regions) and self-gravitating disk (outer region).
In \S\,\ref{sect:observation}, we calculate the thermal emissions due to free-free cooling and nonthermal radiations from the synchrotron and inverse Compton (IC) scattering processes.
In \S\,\ref{sect:discussion}, we discuss the possibility of searching for astronomical transients of the Bondi explosion, especially in radio and $\gamma$-ray bands. Finally, in \S\,\ref{sect:conclusion}, we summarize the main differences and observational signatures of the outflow and cavity in different regions of the AGN disk.

\begin{table*}
\centering
\caption{Studies on the Stars and Compact Objects in AGN Disks.}
\footnotesize
\begin{tabular}{lcccc ccccc ccccc}\hline\hline
\multirow{2}{*}{References} & \multicolumn{4}{c}{Star} && \multicolumn{3}{c}{Compact Object} && \multicolumn{3}{c}{Binary} & \multirow{2}{*}{GRB} &\\
\cline{2-5}\cline{7-9}\cline{11-13}
&Metallicity&Accretion&Evolution&SN&&WD&NS&BH&&Formation&Evolution&Merger&&\\
\hline
\cite{Cheng1999}&\cm&\cm&\cm&&&&&&&\cm&&\cm&\cm&\\
\cite{Baruteau2011}&&&\cm&&&&&&&&\cm&&&\\
\cite{McKernan2012}&&&&&&&&\cm&&&&\cm&&\\
\cite{McKernan2014}&&&&&&&&\cm&&&&\cm&&\\
\cite{Bellovary2016}&&&&&&&&&&&&\cm&&\\
\cite{Stone2017}&&&&&&&&\cm&&&\cm&\cm&&\\ %08 September 2016
\cite{Bartos2017}&&&&&&&&&&&&\cm&&\\ %2017 January 27
\cite{Yang2019}&&&&&&&&&&&&\cm&&\\
\cite{Yi2019}&&&&&&&&&&&&\cm&&\\
\cite{Tagawa2020}&&&&&&&&&&\cm&\cm&&&\\
\cite{Davies2020}&&&&&&&&\cm&&&&&&\\
\cite{McKernan2020}&&&&&&&&&&&&\cm&&\\
\cite{Perna2021EMC}&&&&&&&&&&&&&\cm&\\ % 1.11
\cite{Zhu2021NSmerger}&&&&&&&\cm&\cm&&&&&\cm&\\
\cite{Cantiello2021}&\cm&\cm&&&&&&&&&&&&\\
\cite{PaperI}&&&&&&&&\cm&&&&&&\\   % April 15
\cite{Zhu2021WD}&&&&\cm&&\cm&&&&&&&&\\  % 2021 June 10 
\cite{Jermyn2021}&&&&&&&&\cm&&&&&&\\  % 2021 June 21
\cite{Perna2021AIC}&&&&&&&\cm&&&&&&&\\ % 6.28
\cite{Dittmann2021}&&\cm&&&&&&&&&&&&\\ % 7.26
\cite{PaperII}&&&&&&&&&&\cm&&\cm&&\\ % 7.30
\cite{Grishin2021}&&&&\cm&&&&&&&&&&\\
\cite{Samsing2022}&&&&&&&&&&\cm&&&&\\
\cite{Li2022}&&&&&&&&&&&\cm&&&\\
\cite{Jermyn2022}&&\cm&&&&&&&&&&&&\\
\cite{Yuan2022}&&&&&&&&&&&&&\cm&\\
\cite{Yang2022}&&&\cm&&&&&&&&&&&\\
\cite{McKernan2022}&&&\cm&&&&&&&&&&&\\
\cite{Lazzati2022}&&&&&&&&&&&&&\cm&\\
\cite{Wang2022}&&&&&&&&&&&&&\cm&\\
\cite{LiRixin2022}&&&&&&&&&&&\cm&&&\\
\cite{Ren2022}&&&&&&&\cm&&&&&&&\\
\cite{Fan2023}&\cm&&&&&&&&&&&\cm&&\\
\cite{Li2023Jiaru}&&&&&&&&&&\cm&&&&\\
\cite{Tagawa2023BH}&&&&&&&&\cm&&&&&&\\
\cite{WangMengye2023}&&&&&&&&&&&&\cm&&\\
\cite{Ray2023}&&&&&&&&&&&&&\cm&\\
\cite{Chen2023}&&&&&&&&\cm&&&&&&\\
\cite{Li2023}&&&&\cm&&&&&&&&&&\\
\cite{Zhang2023}&&&&&&\cm&&&&&&&&\\
\cite{Rowan2023}&&&&&&&&&&\cm&&&&\\
\cite{Wang2023}&\cm&\cm&\cm&\cm&&&&&&&&&&\\
\cite{Luo2023}&&&&&&\cm&&&&&&&&\\
\cite{Tagawa2023Shock}&&&&&&&&&&&&\cm&&\\
\cite{Huang2023}&\cm&&&&&&&&&&&&&\\
\cite{Kathirgamaraju2023}&&&&&&&&&&&&\cm&&\\
\cite{Ali-Dib2023}&&&\cm&&&&&&&&&&&\\
\cite{PaperIII}&&&&&&&&\cm&&&\cm&\cm&&\\
\cite{Chen2024}&&&&&&&&&&&&\cm&&\\  % 2024 January 26
\cite{Dittmann2024}&&&&&&&&&&&\cm&&&\\ % 2024 March 15
\cite{Qian2024}&&&&&&&&&&\cm&&&&\\  % 2024 February 15
\cite{Rom2024}&&&&&&&&&&\cm&&&&\\  % 2024 March 14
\cite{Wang2024}&&&&&&&&&&\cm&&&&\\  % 01 February 2024
\cite{ChenLin2024}&&&\cm&&&&&&&&&&&\\ %2024 May 21
\hline
\end{tabular}
\label{tab:ref}
\end{table*}

\section{The model}
\label{sect:model}
\begin{figure*}
\centering
\includegraphics[scale=0.9, trim=430 200 135 110, clip]{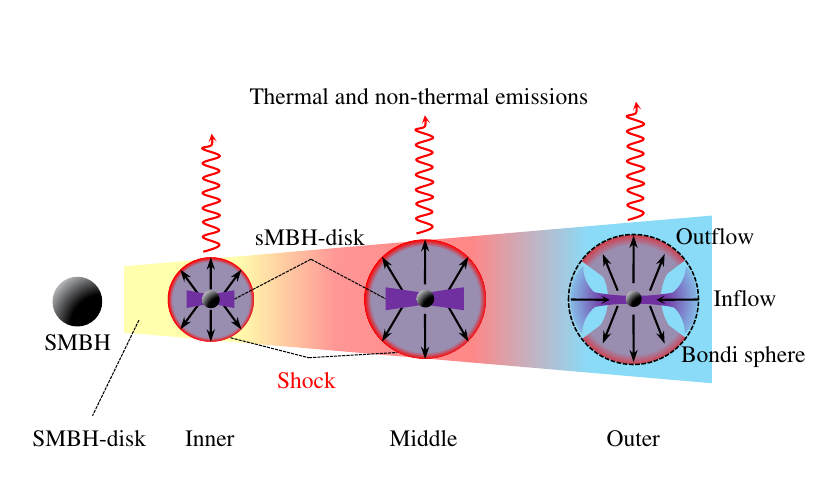}
\caption{\footnotesize Bondi explosion of AMSs embedded in an AGN disk.
AMSs in the inner and middle regions generate powerful outflows and form hot cavities,
whereas for the AMS in the outer region, outflows and inflows coexist.}
\label{fig:cartoon}
\end{figure*}

AMS often undergoes hyper-Eddington accretion (higher than the usual super-Eddington accretion by several orders of magnitude) in the form of Bondi accretion \citep[e.g.,][]{PaperI, PaperII}.
The angular momentum (AM), due to differential rotation, will lead to the formation of a disk  around the sMBH (see Appendix \ref{appdx: sMBHdisk} and Figure \ref{fig:cartoon}) and release large amount of gravitational energy. 
Such a high accretion rate and energy release will give rise to outflows,
which strongly impact the Bondi sphere, generating shocks and forming a cavity.
The cavity is replenished under the pressure of the SMBH disk, leading to episodic accretion onto AMSs.
Also, AMSs will grow heavier through accretion when migrating from the outer region of the SMBH disk to the inner region.
We use the above model frame to study the outflow properties of AMSs with different mass in three regions (Section \ref{sect:application}),
including the SEDs of thermal and nonthermal emissions, light-curve behaviors and its timescales determined by the accretion and outflow timescales (Section \ref{sect:observation}).
We now give detailed general calculations of the AMS model as follows.

We consider AGN disks of SMBHs with mass $\Mp$ and accretion rates $\dot{M}_{\rm p}$.
It is convenient to use the dimensionless accretion rates of $\dotMp = \dot \Mp c^2/L_{\rm Edd,\,p}$,
where $L_{\rm Edd,\,p}= 4\pi G\Mp m_{\rm p}c/\sigT$ is the Eddington luminosity of the SMBH,
$G$ is the gravitational constant,
$m_{\rm p}$ is the proton mass,
$c$ is the speed of light,
$\sigT$ is the Thompson scattering cross section. In this paper, we discuss the case of $\dotMp\sim 1$, namely, SMBH is accreting in the regime of Shakura-Sunyaev model \citep{Shakura1973}.
We consider the simple case that the sMBH is trapped in and corotates with the disk, as shown in Figure \ref{fig:cartoon}.
Then, its dimensionless Bondi accretion rate is described by the Hoyle–Lyttleton–Bondi formulation \citep{Hoyle1939, Bondi1952} and modified in thin disks \citep{Kocsis2011},
\begin{equation}
\label{eq:dotMs}
\begin{array}{llll}
\dotMs & = & \displaystyle \frac{\dot{M}_{\rm s}}{\dot{M}_{\rm Edd, s}}
=  \displaystyle \frac{4\pi G^{2}\Ms^{2}\rhop}{\left({\cs^2+\Delta v^2}\right)^{3/2}\dot{M}_{\rm Edd, s}} 
\displaystyle \times \min\left\{\frac{\Hp}{\RBon},\,1\right\}
\times \min\left\{\frac{\RHill}{\RBon},\,1\right\},
\end{array}
\end{equation}
and the Bondi radius is given by
\begin{equation}
\label{eq:RBon}
\RBon=\frac{G\Ms}{\cs^2+\Delta v^2},
\end{equation}
where $\Ms$ is the mass of the sMBH,
$\rhop$ is the gas density around the sMBH,
$\cs$ is the speed of sound,
$\Hp$ is the half-thickness of the SMBH disk,
$\RHill=(\Ms/3\Mp)^{1/3}R$ is the Hill radius of the sMBH,
$R$ is the locus of the sMBH in the SMBH disk,
$\dot{M}_{\rm Edd, s}= L_{\rm Edd,\,s}/c^2$,
$L_{\rm Edd,\,s}= 4\pi G\Ms m_{\rm p}c/\sigT$ is the Eddington luminosity of the sMBH,
$\Delta v=\OmgK\times\min\{\RBon,\,\RHill\}/2$ is the relative velocity between the sMBH and the gas at the Hill radius or Bondi radius,
$\OmgK=\sqrt{G\Mp/R^3}$ is the Keplerian angular velocity.
The subscripts ``s'' and ``p'' refer to the secondary (i.e., sMBH) and primary BH (i.e., SMBH) of the binary systems.
The second term of Equation (\ref{eq:dotMs}) accounts for the geometry effect of the SMBH disk,
indicating that the Bondi accretion will be suppressed by a factor of $(\Hp/\RBon)$
if the half-thickness of the SMBH disk is smaller than the Bondi radius $\RBon$.
The third term represents the tidal effect of the central SMBH,
suggesting that only the gas located within the Hill radius will undergo accretion onto the sMBH.

Generally, Bondi accretion onto the sMBH in the standard disk is hyper-Eddington due to the dense gas environment \citep[e.g., ][]{PaperI},
resulting in the development of powerful outflows \citep[e.g.][]{Ohsuga2005,Jiang2014,PaperI}.
The outflow power is 
\begin{equation}
\label{eq:Lout}
\Lout=\etaout \dot{M}_{\rm s} c^2,
\end{equation}
\begin{equation}
\label{eq:eta}
\etaout=\eta_{\rm rad}\facc (1-f_{\rm adv}),
\end{equation}
where $\etaout$ is the conversion efficiency of channeling gravitational energy into outflows,
$\eta_{\rm rad}\approx 0.1$ denotes the radiation efficiency,
$f_{\rm adv}\approx 0.9$ is the advection fraction,
and $\facc $ is the fraction of the accretion rate onto the sMBH to the Bondi accretion rate (see Equation (\ref{eq:facc})).
In fact, $f_{\rm adv}$ is quite uncertain in consideration of advection, photon trapping, and outflows\citep[e.g., see the simulations in ][]{Takeuchi2009}.

These powerful outflows will significantly impact the local structures of the SMBH disk,
generating a cavity \citep{PaperI,PaperII,Kimura2021} with a radius denoted as $\Rcav$,
provided that the kinetic energies of the outflows exceed the local dissipation rates of gravitational energy within the SMBH disk \citep{PaperIII},
i.e., $\Lout\gtrsim 4\pi \Rcav^{3}\varepsilon_{+}/3$,
where $\varepsilon_{+}\approx Q_{+}/\Hp$ is the volume dissipation rates of the SMBH disk,
$Q_{+}=3G\Mp\dot{M}_{\rm p}/8\pi R^{3}$ is surface heating rates \citep{Frank2002}, and we disregard the factor of the inner boundary condition of accretion disks. Based on this condition, the maximum of $\Rcav$ can be determined as
\begin{equation}
\label{eq:Rcav}
\Rcav=\left(\frac{3\Lout}{4\pi \varepsilon_{+}}\right)^{1/3}.
\end{equation}

In the outer region of the self-gravitating disks, we demonstrate that the outflows are momentum-driven since the diffusion timescale $\tdiff$ is much shorter than the expansion timescale $\tout$, as shown in Table \ref{tab:SED}.
So the outflow velocity $\vout$ and its expansion timescale $\tout$ satisfy the momentum equation $\Mout \vout / \tout = \Lout/c$ \citep{King2003} and kinematic equation $\Rout = \vout\tout$, respectively,
where $\Rout=\min\{\RBon, \RHill, \Hp, \Rcav\}$ is the outflow radius when the mass falling from SMBH disk is halted,
and $\Mout=4\pi \Rout^3\rhop/3$ is the gas mass within $\Rout$.
Then, we obtain
\begin{equation}
\label{eq:voutMo}
\vout = \left(\frac{\Lout \Rout}{\Mout c}\right)^{1/2},
\end{equation}
\begin{equation}
\label{eq:toutMo}
\tout = \left(\frac{\Mout \Rout c}{\Lout}\right)^{1/2}.
\end{equation}
While in the inner and middle regions, the energy-driven outflows are developed because the diffusion timescale of the photons is much longer than the expansion timescale.
So the outflow velocity $\vout$ and its expansion timescale $\tout$ satisfy the energy equation
$\Lout \tout=\Mout \vout^2/2$,
and the kinematic equation $\Rout = \vout\tout$.
Then, we obtain
\begin{equation}
\label{eq:voutEner}
\vout = \left(\frac{2\Lout \Rout}{\Mout}\right)^{1/3},
\end{equation}
\begin{equation}
\label{eq:toutEner}
\tout = \left(\frac{\Mout \Rout^2}{2\Lout}\right)^{1/3}.
\end{equation}
Here, we approximate $\Mout$ as the total mass of the outflow
given that the initial outflow mass from the sMBH disk $M_{\rm ini}=(1-\facc )\dot{M}_{\rm s}\tout \lesssim \Mout$ during the time interval of $\tout$ and can be ignored (see later calculations in \S\,\ref{sect:application}).
The gravity of the gas within the Bondi radius may enhance the Bondi accretion \citep{Wandel1984, Kocsis2011},
which depends on the ratio of $\Mout/\Ms$.
The accretion onto the sMBH lasts for a timescale of $\tlast=\max\{\tout, \tviss\}$, determined by $\tout$ and the viscosity timescale of the sMBH disk ($\tviss$, see Equation\,\ref{eq:tviss}).
Here, we compare the model with the case of extremely low accretion rates (ADAF disk, $\dotMp\ll1$) discussed in \cite{PaperIII} and point out the difference that the role of external pressure of the SMBH disk is unimportant in our model.
In \S\,\ref{sect:application}, we demonstrate that only a small fraction of the kinetic energy is utilized to work against the external pressure of the SMBH disk.
Because the gas density in the standard disk and self-gravitating disk (see Equations (\ref{eq:diskOut}), (\ref{eq:diskMid}), (\ref{eq:diskInn}), and Figure \ref{fig:SGdisk}) is much higher than that in ADAF disk \citep[$\sim 10^{-14}\rm\,g\,cm^{-3}$,][]{PaperIII} while the temperature is much lower than that in the ADAF disk \citep[$\sim 10^{10}\rm \,K$,][]{Narayan1995ADAF}, the kinematic energy of outflows is much larger than the thermal energy (i.e., the work used to overcome the external pressure) after the outflows rush out of the SMBH disk.
% where all of the kinetic energy is used to overcome the external pressure } and the AMS gravity by detailed calculations
% 

If the velocity of the outflow exceeds the local sound speed,
a shock will emerge and heat the gas in the SMBH disk \citep[e.g.,][]{Blandford1987}.
The temperature of the shocked gas can be determined using the Rankine–Hugoniot  jump conditions,
\begin{equation}
\label{eq:Tsh}
\Tsh=\frac{2(\Gamma_{\rm ad}-1)m_{\rm p}\vout^2}{(1+\Gamma_{\rm ad})^2\kB},
\end{equation}
where the adiabatic index $\Gamma_{\rm ad}=5/3$,
and $\kB$ is the Boltzmann constant.
The heated gas will quickly expand as its gas pressure increases significantly after being shocked.
Once the pressure equilibrium between the shocked gas and SMBH disk is attained,
$\rho_{\rm sh}\kB \Tsh/m_{\rm p}=\rhop \cs^2$,
a cavity is formed \citep{PaperII}.
We then obtain the density of the shocked gas
\begin{equation}
\label{eq:rhosh}
\rho_{\rm sh} =\frac{\rhop \cs^2m_{\rm p}}{\kB \Tsh}.
\end{equation}
The hot gas in the cavity will lose its thermal energy through free-free cooling
within a timescale of
\begin{equation}
\label{eq:dtcool}
\tcoolff =\frac{3\rho_{\rm sh}\kB \Tsh}{2m_{\rm p}j_{\rm ff}},
\end{equation}
where $j_{\rm ff}$ is the free-free emission coefficient.

After cooling,
the cavity will be refilled under the pressure of the SMBH disk on the rejuvenation timescale of \citep{PaperIII}
\begin{equation}
\label{eq:dtrej}
\trej =\frac{\Rcav}{\vturb},
\end{equation}
where $\vturb=\alpha \cs$ is the turbulence velocity.
After the cavity is refilled with cold gas from the SMBH disk,
the above physical process will restart, indicating episodic accretion.
This is very similar to the cases of BH seed growth at high redshift \citep{Wang2006} and accretion onto intermediate mass BHs in dense protogalactic clouds \citep{Milosavljevic2009}.

Apart from the above episodic Bondi explosion, another important feature of the sMBH is the migration from the outer to inner regions due to the torque of the SMBH disk.
Here, we consider two timescales: the viscosity timescale of the SMBH disk and the type I migration timescale given by
\begin{equation}
\label{eq:tvisp}
\tvisp = \frac{R}{\vrp},
\end{equation}
\begin{equation}
\label{eq:tlin}
\tlind = \frac{\Ms R^2 \OmgK}{T_{\rm net}},
\end{equation}
respectively, where $\vrp$ is the radial velocity of the SMBH disk,
$T_{\rm net}=(1.364+0.541\times \gamma_\sigma)(\Ms R\OmgK/\Mp \cs)^2 \Sigp R^4 \OmgK^2$ is the Lindblad torque \citep{Tanaka2002},
$\gamma_\sigma=-d\ln \Sigp/d\ln R$,
and $\Sigp = 2\rhop \Hp$ is the surface density.
The real migration timescale is $t_{\rm mig}=\min (\tvisp, \tlind)$.

\section{AMS in the standard disk and self-gravitating disk}
\label{sect:application}
In this paper, we employ the self-gravitating disk model in the outer region \citep{Sirko2003} and standard model in the middle and inner regions \citep{Shakura1973}.
The disk becomes self-gravity dominated beyond a critical radius $R_{\rm SG}$ determined by the Toomre parameter $Q=\OmgK\cs/\pi G\rhop \Hp=1$ \citep{Toomre1964},
which yields $R_{\rm SG}/\Rg=1.2\times 10^3\,\alpha_{0.1}^{28/45}M_8^{-52/45}\dotMp^{-22/45}$ \citep[e.g.,][]{PaperI}.

\subsection{Outer Region}
\label{sect:outer}

\begin{figure*}
\centering
\includegraphics[scale=0.73, trim=5 0 0 0]{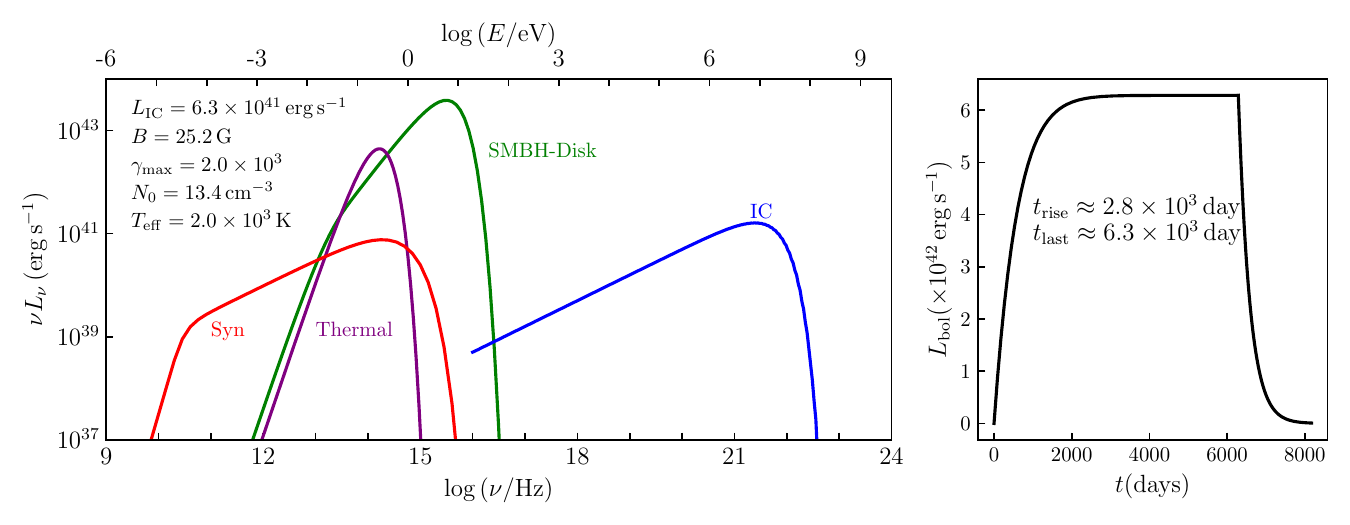}
\caption{\footnotesize {\it Left}: The spectral energy distributions of the powerful outflows developed by the AMS in the outer region of the SMBH disk and the blackbody radiation from the SMBH disk.
The integrated blackbody radiation from the SMBH disk is shown in green.
AGN disk gas is shocked by accretion-powered outflows and emits thermal radiation (purple) with an effective temperature of $\Teff$.
The shock formed from the strong outflows accelerates electrons in AGN disk, emitting synchrotron emissions (red) and generating high-energy photons by IC scattering (blue).
{\it Right}: The thermal light curve of the Bondi explosion, described by Equation (\ref{eq:lc}), which behaves like a Heaviside step function since the inflow and outflow coexist (see Figure \ref{fig:cartoon}).
The rise timescale and the duration are $t_{\rm rise}\approx \teff$ and $\tlast\approx \tout$, respectively.
The related parameters are listed in Table \ref{tab:SED}.}
\label{fig:SEDout}
\end{figure*}

In the outer region beyond $R_{\rm SG}=1.2\times 10^3\,\Rg$, the disk becomes self-gravitating.
As described by Equations (\ref{eq:HSG}), (\ref{eq:rhoSG}), (\ref{eq:csSG}), and (\ref{eq:vpSG}) in Appendix \ref{appdx: SGdisk},
the half-thickness, gas density, sound speed of the SMBH disk, and radial velocity are
\begin{equation}
\label{eq:diskOut}
\left\{
\begin{array}{lcl}
\Hp&=&2.3\times10^{16}\,\alpha_{0.1}^{-1/3} r_5^{3/2}\dotMp^{1/3} M_8^{4/3}\rm\,cm,  \\
[1.0ex]
\rhop&=&9.8\times10^{-15}\,r_5^{-3} M_8^{-2}\rm\,g\,cm^{-3},\\
[1.0ex]
% \Tc&=&4.1\times10^{3}\rm\,K,\\
% [1.0ex]
\cs&=&1.5\times10^6\,\alpha_{0.1}^{-1/3} \dotMp^{1/3} M_8^{1/3}\rm \, cm\,s^{-1},
\\
[1.0ex]
\vrp &=&3.4\times10^{3}\,\alpha_{0.1}^{1/3} r_5^{1/2}\dotMp^{2/3} M_8^{2/3}\rm\,cm\,s^{-1},
\end{array}
\right.
\end{equation}
where $\alpha_{0.1}=\alpha/0.1$ is the viscosity parameter,
$r_5=R/10^5\,\Rg$,
and $M_8= \Mp/10^8M_{\odot}$ denotes the mass of the central SMBH. 
By substituting the disk parameters into Equations (\ref{eq:dotMs}) and (\ref{eq:RBon}), we can estimate the dimensionless Bondi accretion rate and Bondi radius as follows:
\begin{equation}
\label{eq:dotMsOut}
\dotMs
=5.0\times 10^5\,\alpha_{0.1} r_5^{-3}\dotMp^{-1} M_8^{-2}\qq,
\end{equation}
\begin{equation}
\RBon=6.2\times 10^{15}\,\alpha_{0.1}^{2/3} \dotMp^{-2/3} M_8^{1/3}\qq\,{\rm cm},
\end{equation}
where $\qq=q/10^{-6}$, $q=\Ms/\Mp=10^{-6}m_2/M_8$, and $m_2=\Ms/10^2\,\Msun$.
First, the Bondi radius is much smaller than the disk half-thickness,
so the geometry effect can be ignored.
Second, the Hill radius of the sMBH is $\RHill=(\Ms/3\Mp)^{1/3}R
\approx 1.0\times 10^{16}\,{\rm cm}$,
much larger than the Bondi radius, so the tidal effect of the SMBH is also marginal.
Third, the differential velocity $\Delta v=\OmgK\RBon/2=2.0\times 10^5\,\rm cm\,s^{-1}$, is about 1 order of magnitude lower than the sound speed,
indicating that the modification to the spherical accretion can also be ignored.
Finally, the sphere mass within the Bondi radius is $\MBon=4\pi \RBon^3\rhop/3=5.0\,M_{\odot}$,
much smaller than the sMBH mass,
implying that its gravity only has a marginal effect on enhancing the accretion process.

By replacing size $\Delta R$ in Equation (\ref{eq:Xout}) with $\RBon$, we obtain the outer radius of the sMBH disk as
\begin{equation}
\label{eq:XoutOut}
\Xout=\frac{\RBon^4\Mp}{4 R^3\Ms}
=1.2\times 10^{14}\,\alpha_{0.1}^{8/3} r_5^{-3}\dotMp^{-8/3} M_8^{-5/3}\qq^{3}\,\rm cm,
\end{equation}
corresponding to $7.9\times10^6\,\rg$,
where $\rg=G\Ms/c^2$ is the gravitational radius of the sMBH.
This gives an accretion fraction $\facc\approx 8.7\times 10^{-4}$.
However, the formation of the sMBH disk can be quenched by the turbulence with a size scale of $l_{\rm turb} \sim \alpha \cs/\OmgK \sim \alpha \Hp \sim 2.3\times 10^{15}\,\rm cm$, comparable with $\RBon$.
The chaotic turbulence leads to a low AM material accretion onto the sMBH \citep[][]{Chen2023} and gives rise to a small-sized sMBH disk as indicated by its AM expression $\Delta \ell_{\rm s}=\sqrt{\Xout G\Ms}$.
The simulations of Bondi-Hoyle accretion in a turbulent medium show that turbulence reduces the accretion rate by a factor of a few, possibly dependent on the magnetic field and the Mach number of the accreting gas \citep[e.g.,][]{Lee2014, Burleigh2017}.
In consideration of the above uncertainty of the turbulence, we set $\facc \sim 0.1$ and define a parameterized conversion efficiency $\etaout$.
With Equation (\ref{eq:eta}), we obtain the conversion efficiency $\etaout\approx 10^{-3}$.
Based on Equations (\ref{eq:Lout}) and (\ref{eq:dotMsOut}), the sMBH experiences a hyper-Eddington accretion process and forms powerful outflow with
\begin{equation}
\label{eq:LoutOut}
\Lout = 6.3\times10^{42}\,\etatwo\alpha_{0.1} r_5^{-3}\dotMp^{-1}M_8^{-1} \qq^{2}\,\rm erg\,s^{-1},
\end{equation}
where $\etatwo=\etaout/10^{-3}$.

With the condition that the power of the outflow exceeds the local heat production rate of the AGN disk, the cavity radius is given by Equations (\ref{eq:Rcav}) and (\ref{eq:LoutOut}) as $\Rcav=1.7\times 10^{18}\,{\rm cm}$,
much larger than the disk half-thickness.
This indicates that the outflow can rush out of the disk and significantly change the local disk structure.
In this case, the maximum cavity radius is
\begin{equation}
\label{eq:RcavOut}
\Rcav=\Hp=2.3\times10^{16}\,\alpha_{0.1}^{-1/3} r_5^{3/2}\dotMp^{1/3} M_8^{4/3}\rm\,cm.
\end{equation}
The gas mass within the cavity is given by $\Mcav=4\pi\Rcav^3\rhop/3=2.4\times10^2\,\Msun$.
When the accretion is halted, the outflow radius is $\Rout=\min\{\RBon, \RHill, \Hp, \Rcav\}=\RBon=6.2\times 10^{15}\rm\,cm$ and the outflow mass is $\Mout=4\pi\Rout^3\rhop/3=5.0\,\Msun$.
Based on Equations (\ref{eq:voutMo}), (\ref{eq:toutMo}), and (\ref{eq:LoutOut}), we derive the velocity of the outflow and its expansion timescale as
\begin{equation}
\label{eq:voutOut}
\vout=1.2\times 10^7\,\etatwo^{1/2}\alpha_{0.1}^{-1/6}  \dotMp^{1/6}M_8^{1/6}\,\rm cm\,s^{-1},
\end{equation}
\begin{equation}
\label{eq:toutOut}
\tout=17.2\,\etatwo^{-1/2}\alpha_{0.1}^{5/6}
\dotMp^{-5/6} M_8^{1/6} \qq\,\yr.
\end{equation}
The energy of outflows is $\Eout=\Lout \tout=3.4\times 10^{51}\,\rm erg$.
We calculate the energy required to work against the external pressure and the AMS gravity.
First, the energy used to overcome the pressure of the SMBH disk is $\Ep=4\pi P\Rout^3/3=1.0\times 10^{48}\,\rm erg$, much smaller than $\Eout$, indicating that it is easy to overcome the disk pressure and the effect can be ignored.
Second, the energy used to overcome the gravity of the AMS is $\Eg=G (\Mout+\Ms)\Mout/\Rout=2.2\times 10^{46}\,\rm erg$.
These two considerations verify that almost all of the dissipated energy of the sMBH disk is transformed into the kinetic energy of the outflow $\Eout$.
We also calculate the initial mass of outflows $M_{\rm ini}\approx (1-\facc)\dot{M}_{\rm s}\tout=1.7\,M_{\odot}\lesssim \Mout$,
validating the previous approximation of ignoring it in Equations (\ref{eq:voutOut}) and (\ref{eq:toutOut}).

The velocity of the outflow is about 1 order of magnitude larger than the sound speed, inevitably leading to shock formation and the generation of relativistic electrons and nonthermal radiation, as shown in Figure \ref{fig:SEDout}.
Meanwhile, based on Equations (\ref{eq:Tsh}) and (\ref{eq:voutOut}), the gas will be heated to a high temperature of $\Tsh=3.0\times10^5\,\rm K$ by the strong shock.
Utilizing the pressure equilibrium condition between the shocked gas and SMBH disk in Equation (\ref{eq:rhosh}),
we derive the density of shocked gas $\rho_{\rm sh} = 8.5\times 10^{-16}\rm \,cm^{-3},$
much lower than the density of the SMBH disk gas.
After the outflow rushes out of the SMBH disk,
it will form a cavity of shocked gas with high temperature and low density.
The hot gas loses its thermal energy mainly by free-free cooling with a cooling timescale of $\tcoolff =2.2\,\rm d$ given by Equation (\ref{eq:dtcool}),
much shorter than the cavity expansion timescale $\tout$.

After the hot gas in the cavity is cooled quickly by free-free emission,
the cavity is refilled with the surrounding cold gas turbulence.
Based on Equations (\ref{eq:dtrej}) and (\ref{eq:RcavOut}),
the cavity can be refilled in a rejuvenation timescale of
\begin{equation}
\label{eq:dtrejValueOuter}
\trej =4.9\times 10^3\,\alpha_{0.1}^{-1} r_5^{3/2}M_8\,\yr.
\end{equation}
The rejuvenation is much slower than the cavity formation and leads to episodic accretion.
Although the sMBH disk size is much smaller than the result of Equation (\ref{eq:XoutOut}),
we calculate the lower limit of the radial velocity of the sMBH disk and the upper limit of the viscosity timescale based on Equations (\ref{eq:vrs}), (\ref{eq:tviss}), and (\ref{eq:XoutOut}),
\begin{equation}
\vrs
=4.3\times 10^5\,\alpha_{0.1}^{-1/3} r_5^{3/2}\dotMp^{4/3} M_8^{4/3} \qq^{-1}\,\rm cm\,s^{-1},
\end{equation}
\begin{equation}
\tviss
=8.7\,\alpha_{0.1}^{3} r_5^{-9/2}\dotMp^{-4} M_8^{-3} \qq^{4}\,\rm yr,
\end{equation}
which is smaller than the expansion timescale $\tout$.
This means that the inflow and outflow coexist, as shown in Figure \ref{fig:cartoon}.
The sMBH disk gas will keep accreting onto the sMBH for a timescale of $\tlast=\max\{\tout, \tviss\}=\tout$,
until the outflow expands to the Bondi radius.
The corresponding duty cycle is $\delta = \tlast/(\tlast+\trej)=3.5\times10^{-3}$.
The average mass growth timescale of the sMBH is 
$\tgrow=\Ms/\facc \delta\dotMs\dot{M}_{\rm Edd, s}=2.6\times 10^6 \,\yr$.
Using Equations (\ref{eq:tvisp}) and (\ref{eq:tlin}), we obtain the viscosity timescale of the SMBH disk $\tvisp=1.4\times 10^7\rm \, yr$, and type I migration timescale $\tlind = 1.1\times 10^7\rm \, yr$, respectively,
where $\gamma_\sigma=-d\ln \Sigp/d\ln R=3/2$ is used.
The real migration timescale of the sMBH is $\tmig=\tlind$.
For an exponentially growing case, the sMBH mass can grow up to $\Ms^\prime= \exp (\tlind/\tgrow)\Ms=69.0\,\Ms$.

\subsection{Middle Region}
\label{sect:middle}

\begin{figure*}
\centering
\includegraphics[scale=0.73, trim=5 0 0 0]{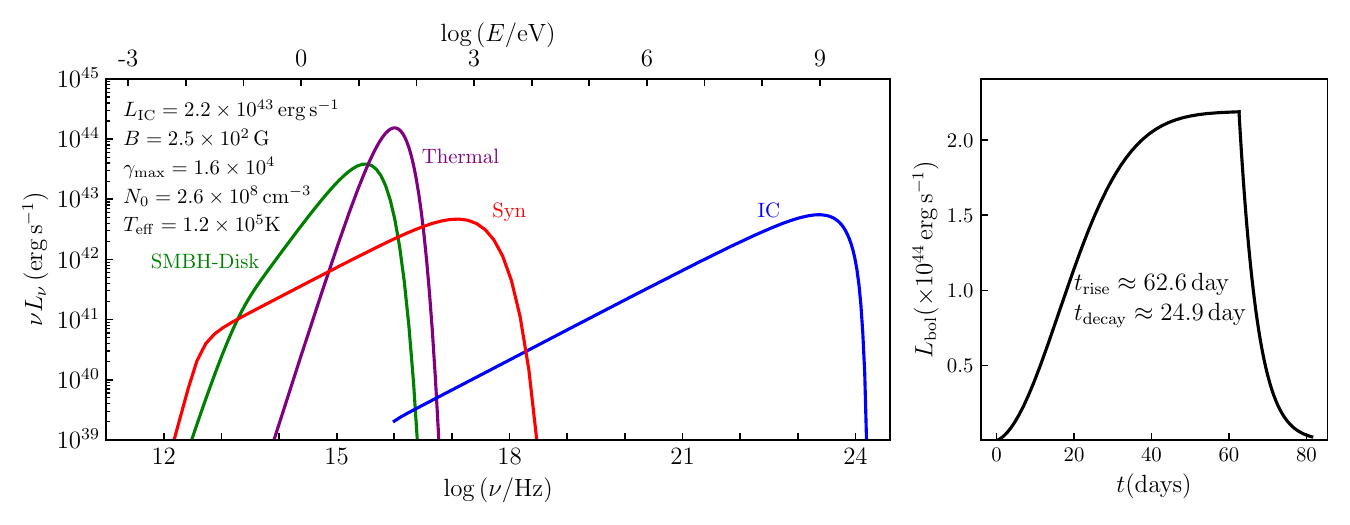}
\caption{\footnotesize The same as the Figure \ref{fig:SEDout} but for the middle region in the SMBH disk.
{\it Left: }The blackbody radiation of the SMBH disk is shown in green.
The shocked SMBH disk gas emits thermal radiation (purple) with an effective temperature of $\Teff$.
Nonthermal electrons accelerated through shocks emit synchrotron emissions (red) and generate high-energy photons by IC scattering (blue).
{\it Right}: The thermal light curve experiences slow rise and rapid decay with timescales of $t_{\rm rise}\approx\tlast$ and $t_{\rm decay}\approx\teff$, respectively.}
% The rise timescale is and the decay timescale is .}
\label{fig:SEDmid}
\end{figure*}

In the middle region of the standard disk, gas pressure prevails over radiation pressure, and the electron scattering optical depth is significantly larger than the absorption.
According to the standard model, the radius of the middle region is $3.0\times10^2\, (\alpha_{0.1}M_8)^{2/21}\dotMp^{16/21}<R/\Rg \le 5.1\times10^3\,\dotMp^{2/3}$
\citep[][]{Svensson1994},
and the half-thickness, density, midplane temperature, sound speed, and radial velocity of the SMBH disk are \citep{Kato2008}
\begin{equation}
\label{eq:diskMid}
\left\{
\begin{array}{lcl}
\Hp&=&3.6\times10^{13}\,\alpha_{0.1}^{-1/10}r_3^{21/20}\dotMp^{1/5} M_8^{9/10}\rm\,cm,  \\[1.0ex]
\rhop&=&3.4\times10^{-9}\,\alpha_{0.1}^{-7/10}r_3^{-33/20}\dotMp^{2/5} M_8^{-7/10}\rm\,g\,cm^{-3},\\
[1.0ex]
\Tc&=&3.2\times10^{4}\,\alpha_{0.1}^{-1/5}r_3^{-9/10}\dotMp^{2/5} M_8^{-1/5}\rm\,K,\\
[1.0ex]
\cs&=&2.3\times10^6\,\alpha_{0.1}^{-1/10}r_3^{-9/20}\dotMp^{1/5} M_8^{-1/10}\rm \, cm\,s^{-1},
\\
[1.0ex]
\vrp&=&5.9\times10^{2}\,\alpha_{0.1}^{4/5}r_3^{-2/5}\dotMp^{2/5} M_8^{-1/5}\rm\,cm\,s^{-1},
\end{array}
\right.
\end{equation}
where $r_3=R/10^3\,\Rg$.

As discussed in \S\,\ref{sect:outer}, the sMBH grows heavier due to the episodic accretion during the migration.
We, therefore, consider an sMBH with $10^3\,\Msun$ mass in the middle region.
By substituting the disk parameters into Equations (\ref{eq:dotMs}) and (\ref{eq:RBon}),
we can estimate the dimensionless modified Bondi accretion rate and Bondi radius as follows:
\begin{equation}
\label{eq:dotMsMid}
\dotMs=1.7\times 10^7\,\alpha_{0.1}^{-4/5}r_3^{-1/10}\dotMp^{3/5} M_8^{1/5} \qqsix^{-1/3},
\end{equation}
\begin{equation}
\RBon=2.7\times 10^{15}\,r_3 M_8 \qqsix^{1/3}\,{\rm cm},
\end{equation}
where $\qqsix=q/10^{-5}$, $q=\Ms/\Mp=10^{-5}m_3/M_8$, and $m_3=\Ms/10^3\,\Msun$.
First, the Bondi radius is significantly larger than the SMBH disk half-thickness, leading to a suppression of the Bondi accretion rate by about 2 orders of magnitude.
Second, the Hill radius, given by $\RHill=(\Ms/3\Mp)^{1/3}R
\approx 2.2\times 10^{14}\,{\rm cm}$,
is also smaller than the Bondi radius, resulting in a suppression of the Bondi accretion rate by 1 order of magnitude.
Third, we calculate the differential velocity $\Delta v=\OmgK\RHill/2=7.1\times 10^6\,\rm cm\,s^{-1}$, 
which is approximately three times larger than the sound speed.
This suggests that the accretion deviates from spherical Bondi accretion.
So we ignore the sound speed when calculating $\dotMs$ and $\RBon$.
Finally, the gas mass within the Hill radius is $\MBon=2\pi\Hp\RHill^2\rhop/3=6.4\,\Msun$,
much smaller than the sMBH mass,
indicating that the enhancement effect on the accretion rate from the gas self-gravity within the Bondi sphere is marginal.

By substituting $\Delta R$ in Equation (\ref{eq:Xout}) with $\RHill$, we can determine the outer radius of the sMBH disk as
\begin{equation}
\label{eq:XoutMid}
\begin{array}{llll}
\Xout & = &\displaystyle \frac{\RHill^4\Mp}{4 R^3\Ms} 
= 1.8\times 10^{13}\, r_3 M_8 \qqsix^{1/3}\,\rm cm.
\end{array}
\end{equation}
This corresponds to $1.2\times 10^5\,\rg$,
resulting in an accretion fraction of $\facc\approx 6.9\times 10^{-3}$.
Utilizing Equation (\ref{eq:eta}), we derive a conversion efficiency of $\etaout\approx 6.9\times 10^{-5}$.
Here, we would like to point out that the turbulence scale $l_{\rm turb}\sim\alpha \Hp \sim 0.1 \Hp$ is much smaller than the size of accretion region $\min\{\RBon, \RHill, \Hp\}=\Hp$,
so the turbulence will not affect the formation of sMBH disk and can be ignored.
Based on Equation (\ref{eq:Lout}) and (\ref{eq:dotMsMid}), the sMBH forms outflows with a power given by
\begin{equation}
\label{eq:LoutMid}
\Lout
=2.2\times10^{44}\,\etafour\alpha_{0.1}^{-4/5}r_3^{-1/10}\dotMp^{3/5}M_8^{6/5} \qqsix^{2/3}\,\rm erg\,s^{-1},
\end{equation}
where $\etafour=\etaout/10^{-4}$.

Based on Equations (\ref{eq:Rcav}) and (\ref{eq:LoutMid}), the powerful outflows interact with the SMBH disk and form a cavity with a radius of $R_{\rm cav}=6.5\times 10^{15}\,\rm cm$,
which is much larger than the half-thickness of the SMBH disk.
This indicates that the outflow can rapidly rush out of the SMBH disk and significantly alter its local structure.
Therefore, the maximum cavity radius is
\begin{equation}
\label{eq:RcavMid}
R_{\rm cav}=\Hp=3.6\times10^{13}\,\alpha_{0.1}^{-1/10}r_3^{21/20}\dotMp^{1/5} M_8^{9/10}\rm\,cm.
\end{equation}
The gas mass within the cavity is $\Mcav=4\pi\Rcav^3\rhop/3=0.34\,\Msun$.
When the accretion is halted, the outflow radius is $\Rout=\min\{\RBon, \RHill, \Hp, \Rcav\}=\Hp=3.6\times 10^{13}\rm\,cm$ and the outflow mass is $\Mout=4\pi \Rout^3\rhop/3=0.34\,\Msun$.
With Equations (\ref{eq:voutEner}), (\ref{eq:toutEner}), and (\ref{eq:LoutMid}), the velocity of the outflow and its expansion timescale are derived as
\begin{equation}
\label{eq:voutMid}
\vout=2.9\times 10^8\,\etafour^{1/3} \alpha_{0.1}^{1/30} r_3^{-11/60}\dotMp^{-1/15}M_8^{1/30}\qqsix^{2/9} \,\rm cm\,s^{-1},
\end{equation}
\begin{equation}
\label{eq:toutMid}
\tout=1.5\,\etafour^{-1/3}\alpha_{0.1}^{-2/15} r_3^{37/30}\dotMp^{4/15} M_8^{13/15} \qqsix^{-2/9}\,\rm d.
\end{equation}
The energy of outflows is $\Eout=\Lout \tout=2.8\times 10^{49}\,\rm erg$.
We calculate the energy required to work against the external pressure and the AMS gravity.
First, the energy used to overcome the pressure of the SMBH disk is $\Ep=4\pi P\Rout^3/3=3.7\times 10^{45}\,\rm erg$.
Second, the energy used to overcome the gravity of the AMS is $\Eg=G \Mout\Ms/\Rout=2.5\times 10^{48}\,\rm erg$.
These two kinds of energies are much smaller than the outflow energy $\Eout$, indicating that the influence of SMBH disk pressure and AMS gravity on the outflow expansion could be ignored.
We also calculate the initial mass of outflows $M_{\rm ini}\approx (1-\facc)\dot{M}_{\rm s}\tout=0.15\,M_{\odot}\lesssim \Mout$,
which verifies the validity of the former approximation of ignoring it in Equations (\ref{eq:voutMid}) and (\ref{eq:toutMid}).

The outflow velocity is about 2 orders of magnitude larger than the local sound speed,
which gives rise to powerful shocks when the outflows collide with the SMBH disk gas.
The shocks would accelerate the electrons and lead to nonthermal emissions, as can be seen in Figure \ref{fig:SEDmid}.
Meanwhile, the shock will heat the SMBH disk gas to an extremely high temperature $\Tsh=1.9\times10^8\,\rm K$ given by Equations (\ref{eq:Tsh}) and (\ref{eq:voutMid}).
Utilizing the pressure equilibrium condition between the shocked gas and SMBH disk based on Equation (\ref{eq:rhosh}),
we derive the density of the shocked gas, $\rho_{\rm sh} = 1.2 \times 10^{-12}\,\rm g\, cm^{-3}$,
about 3 orders of magnitude lower than the density of the SMBH disk gas. 
With Equation (\ref{eq:dtcool}), the cooling timescale of the shocked gas is given by $\tcoolff=3.4\times 10^3\,\rm s$,
much shorter than the cavity expansion timescale,
indicating that the free-free cooling mechanism is very efficient.

After the hot gas in the cavity is rapidly cooled, the cavity, based on Equations (\ref{eq:dtrej}) and (\ref{eq:RcavMid}), will be refilled in a rejuvenation timescale of
\begin{equation}
\trej =4.9\,\alpha_{0.1}^{-1}r_{3}^{3/2}M_8\,\yr.
\end{equation}
This rejuvenation is much slower than the cavity expansion.
Based on Equations (\ref{eq:vrs}), (\ref{eq:tviss}), and (\ref{eq:XoutMid}),
we can derive the radial velocity and the viscosity timescale of the sMBH disk,
\begin{equation}
\vrs=3.4\times 10^6\,\alpha_{0.1} r_3^{-1/2} \qqsix^{1/3}\,\rm cm\,s^{-1},
\end{equation}
\begin{equation}
\tviss=62.6\,\alpha_{0.1}^{-1}r_3^{3/2}M_8\,\rm d,
% {\color{red}how~ to ~get ~it?}.
\end{equation}
The viscosity timescale is much longer than the expansion timescale $\tout$.
This means that after the outflows rush out of the SMBH disk, the accretion onto AMS still continues and lasts for a timescale of $\tlast=\max\{\tout, \tviss\}=\tviss$.
The duty cycle of the accretion process is $\delta = \tlast/(\tlast+\trej)=3.4\times 10^{-2}$.
The average net mass accretion rate of the sMBH is 
$\dot{M}_{\rm ave} = \facc \delta\dotMs\dot{M}_{\rm Edd, s} =4.2\times 10^3\,\dot{M}_{\rm Edd, s}$.
The average mass growing timescale of the sMBH is 
$\tgrow=\Ms/\dot{M}_{\rm ave}=1.1\times 10^5 \,\yr$.
Using Equations (\ref{eq:tvisp}) and (\ref{eq:tlin}), we obtain the viscosity timescale of the SMBH disk, $\tvisp=7.9\times 10^5\rm \, yr$, and type I migration timescale, $\tlind = 6.4\times 10^2\rm \, yr$, respectively,
where $\gamma_\sigma=-d\ln \Sigp/d\ln R=3/5$ is used.
The real migration timescale of the sMBH is $\tmig=\tlind \ll \tgrow$, indicating that the sMBH can hardly grow up before migrating to the inner region.

\subsection{Inner Region}

\begin{figure*}
\centering
\includegraphics[scale=0.73, trim=5 0 0 0]{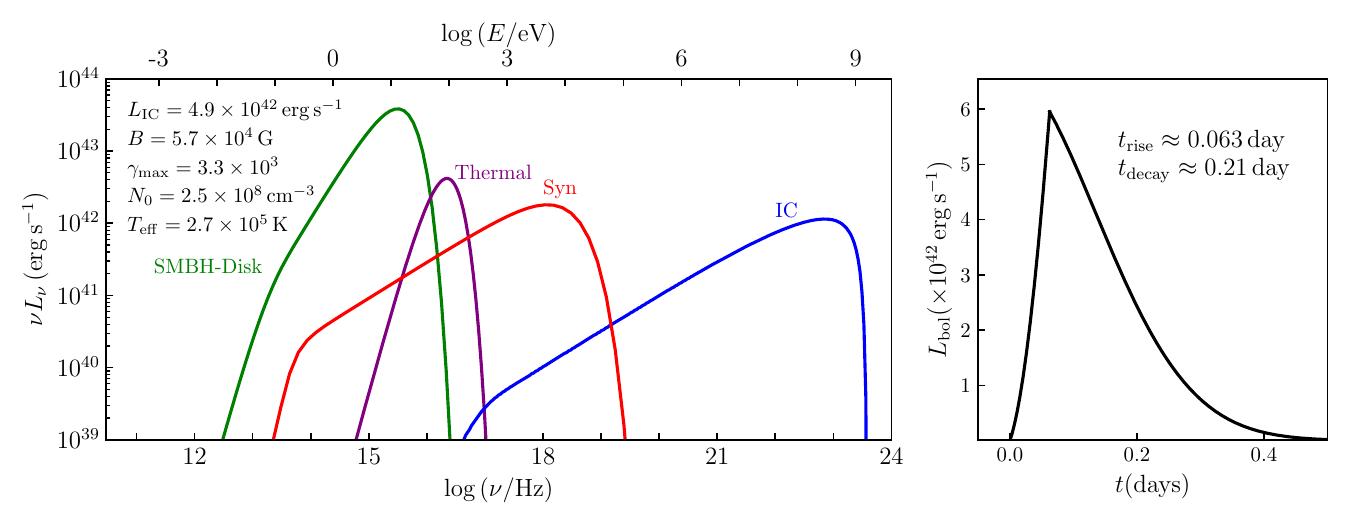}
\caption{\footnotesize
The same as Figure \ref{fig:SEDmid} but for the outer region in the SMBH disk.
{\it Left: }The blackbody radiation of the SMBH disk is shown in green.
The shocked SMBH disk gas emits thermal radiation (purple) with an effective temperature of $\Teff$.
Nonthermal electrons accelerated through shocks emit synchrotron emissions (red) and generating high-energy photons by IC scattering (blue).
{\it Right}: The thermal light curve increases rapidly within a timescale of $t_{\rm rise}\approx\tlast$ and decreases in a timescale of $t_{\rm decay}\approx\teff$, in contrast to the middle region case.}
% The rise timescale is  and the decay timescale is .}
\label{fig:SEDinner}
\end{figure*}

In the inner region of a standard disk, radiation pressure prevails over gas pressure,
and the electron scattering depth is significantly greater than absorption.
According to the standard model, the inner region radius is $R/\Rg \le 3.0\times10^2\, (\alpha_{0.1}M_8)^{2/21}\dotMp^{16/21}$ \citep[Equation (13) or (25) in][]{Svensson1994},
and the half-thickness, density, midplane temperature, sound speed, and radial velocity of the SMBH disk are \citep{Kato2008}
\begin{equation}
\label{eq:diskInn}
\left\{
\begin{array}{lcl}
\Hp&=&1.2\times10^{12}\,M_8\dotMp\rm\,cm,  \\
[1.0ex]
\rhop&=&2.0\times10^{-8}\,\alpha_{0.1}^{-1}M_8^{-1}\dotMp^{-2}r_1^{3/2}\rm\,g\,cm^{-3},\\
[1.0ex]
\Tc&=&4.8\times10^{5}\,\alpha_{0.1}^{-1/4}M_8^{-1/4}r_1^{-3/8}\rm\,K,\\
[1.0ex]
\cs&=&8.0\times10^7\,\dotMp r_1^{-3/2}\rm \, cm\,s^{-1},\\
[1.0ex]
\vrp&=&3.1\times10^{5}\,\alpha_{0.1}\dotMp^{2}r_1^{-5/2}\rm\,cm\,s^{-1},
\end{array}
\right.
\end{equation}
where $r_1 = R/10\,\Rg$ is the radius of the disk from the SMBH.

As discussed in \S\,\ref{sect:middle}, the sMBH hardly grows, due to its quick migration.
We, therefore, still consider an sMBH with $10^3\,\Msun$ mass in the inner region.
Substituting the disk parameters into Equations (\ref{eq:dotMs}) and (\ref{eq:RBon}), the Bondi accretion rate and Bondi radius can be estimated as
\begin{equation}
\label{eq:dotMsInn}
\dotMs=3.9\times 10^{5}\,\alpha_{0.1}^{-1} r_1 \qqfive^{-2/3},
\end{equation}
\begin{equation}
\label{eq:RBonInn}
\RBon=2.1\times 10^{13}\,M_8\dotMp^{-2}r_1^3 \qqfive\,{\rm cm}.
\end{equation}
To derive Equations (\ref{eq:dotMsInn}) and (\ref{eq:RBonInn}), four conditions are considered.
First, the Hill radius $\RHill=(\Ms/3\Mp)^{1/3}R
\approx 2.2\times 10^{12}\,{\rm cm}$,
which is about 1 order of magnitude smaller than the Bondi radius.
This indicates that the tidal effect of SMBH must be considered.
Second, the Bondi radius is about 1 order of magnitude larger than the SMBH disk half-thickness,
which means that the SMBH disk geometry suppresses the accretion rate.
Third, the differential velocity between the gas in the Hill radius and the sMBH is expressed as
$\Delta v=\OmgK\RHill/2=7.1\times10^7\,\rm cm\,s^{-1}$,
which is comparable with the sound speed and not considered in this case just for a rough estimation.
Finally, the gas mass within the Hill radius is $\MBon=2\pi\Hp\RHill^2\rhop/3=1.3\times 10^{-4}\,\Msun\ll \Ms$,
indicating that the enhancement effect from the self-gravity of Bondi sphere gas is marginal.

The outer radius of sMBH disk $\Xout$ can be estimated through the conservation of AM based on Equation (\ref{eq:Xout}).
By setting $\Delta R= \RHill$,
we derive 
\begin{equation}
\label{eq:XoutInn}
\Xout=\frac{\RHill^4\Mp}{4 R^3\Ms}
=1.8\times 10^{11}\,r_1 M_8 \qqfive^{1/3}\,\rm cm.
\end{equation}
This corresponds to $1.2\times 10^3\,\rg$.
With Equation (\ref{eq:facc}), the accretion fraction $\facc\approx 6.9\times 10^{-2}$ is obtained.
With Equation (\ref{eq:eta}), the conversion efficiency $\etaout\approx 6.9\times 10^{-4}$ is derived.
Like the middle region case, the turbulence scale $l_{\rm turb}\sim\alpha \Hp \sim 0.1 \Hp$ is much smaller than the size of accretion region $\min\{\RBon, \RHill, \Hp\}=\Hp$,
so the uncertainty of $\facc$ introduced by turbulence can be ignored.
Based on Equation (\ref{eq:Lout}) and (\ref{eq:dotMsInn}), the power of the outflow is given by
\begin{equation}
\label{eq:LoutInn}
\Lout\approx 4.9\times10^{43} \etathree\alpha_{0.1}^{-1} r_1 M_8 \qqfive^{1/3}\,\rm erg\,s^{-1}, 
\end{equation}
where $\etathree=\etaout/10^{-3}$.

Based on Equations (\ref{eq:Rcav}) and (\ref{eq:LoutInn}), the cavity radius is derived $\Rcav=1.3\times 10^{13}\,{\rm cm}$.
The derived cavity radius is much larger than the SMBH disk half-thickness,
which indicates that the outflow can significantly change the local disk structure.
In this case, the maximum cavity radius is
\begin{equation}
\label{eq:RcavInn}
\Rcav=\Hp=1.2\times10^{12}\,M_8\dotMp\rm\,cm.
\end{equation}
The gas mass within the cavity volume is $\Mcav=4\pi \Rcav^3 \rhop/3=8.0\times10^{-5}\,M_{\odot}$.
When the accretion is halted, the outflow radius is $\Rout=\min\{\RBon, \RHill, \Hp, \Rcav\}=\Hp=1.2\times 10^{12}\rm\,cm$ and the outflow mass is $\Mout=4\pi\Rout^3\rhop/3=8.0\times10^{-5}\,\Msun$.
Using equations (\ref{eq:voutEner}), (\ref{eq:toutEner}), and (\ref{eq:LoutInn}), we obtain the velocity of the outflow and the cavity expansion time
\begin{equation}
\label{eq:voutInn}
\vout=9.1\times 10^8\,\etathree^{1/3} r_1^{-1/6} \qqfive^{1/9} \,\rm cm\,s^{-1},
\end{equation}
\begin{equation}
\label{eq:toutInn}
\tout=1.4\times 10^3\,\etathree^{-1/3} r_1^{1/6}\dotMp M_8 \qqfive^{-1/9}\,\rm s.
\end{equation}
The energy of outflows is $\Eout=\Lout \tout=6.6\times 10^{46}\,\rm erg$.
We calculate the energy required to work against the external pressure and the AMS gravity.
First, the energy used to overcome the pressure of the SMBH disk is $\Ep=4\pi P\Rout^3/3=1.0\times 10^{45}\,\rm erg$.
Second, the energy used to overcome the gravity of the AMS is $\Eg=G \Mout\Ms/\Rout=1.7\times 10^{46}\,\rm erg$.
These two kinds of energies are much smaller than the outflow energy $\Eout$, indicating that the influence of SMBH disk pressure and AMS gravity on the outflow expansion could be ignored.
We also calculate the initial mass of outflows $M_{\rm ini}\approx (1-\facc)\dot{M}_{\rm s}\tout=3.4\times 10^{-5}\,M_{\odot} \lesssim \Mout$,
which verifies the validity of former approximation of ignoring it in Equations (\ref{eq:voutInn}) and (\ref{eq:toutInn}).

The velocity of outflow is about 1 order of magnitude larger than the sound speed,
which inevitably leads to the formation of shock
when the outflow collides with the surrounding gas medium.
The powerful shock accelerates the thermal electrons to relativistic state,
which gives rise to nonthermal emissions due to the synchrotron radiation and IC scattering processes, as can be seen in Figure \ref{fig:SEDinner}.
Meanwhile, the shock will also heat the SMBH disk gas with an extremely high temperature of $\Tsh=1.9\times 10^9\,\rm K$, given by Equations (\ref{eq:Tsh}) and (\ref{eq:voutInn}).
This can lead to the generation of relativistic jet if the sMBH is rapidly spinning \citep{Rees1982}.
Using the pressure equilibrium condition between the shocked gas and SMBH disk based on Equation (\ref{eq:rhosh}),
we derive the density of shocked gas $\rho_{\rm sh} = 8.0 \times 10^{-10}\,\rm g\, cm^{-3}$,
much lower than the density of the SMBH disk gas.
With Equation (\ref{eq:dtcool}),
the cooling timescale of the shocked gas can be derived $\tcoolff
=16.4\,\rm s$,
much shorter than the cavity expansion timescale,
which implies that the free-free cooling mechanism is very efficient.

After the hot gas in the cavity is cooled quickly,
the cavity, based on Equations (\ref{eq:dtrej}) and (\ref{eq:RcavInn}), will be refilled in a rejuvenation timescale of
\begin{equation}
\label{eq:dtrejValue}
\trej=1.8\,\alpha_{0.1}^{-1}r_{1}^{3/2}M_8\,{\rm d},
\end{equation}
much longer than the cavity formation timescale.
Based on Equations (\ref{eq:vrs}), (\ref{eq:tviss}), and (\ref{eq:XoutInn}), the radial velocity and the viscosity timescale of the sMBH disk are derived,
\begin{equation}
\vrs=3.4\times 10^7\,\alpha_{0.1} r_1^{-1/2} \qqfive^{1/3}\,\rm cm\,s^{-1},
\end{equation}
\begin{equation}
\tviss=1.5\,\alpha_{0.1}^{-1}r_1^{3/2} M_8\,\rm hr.
\end{equation}
The viscosity timescale is much longer than the expansion timescale $\tout$.
This means that after the outflows rush out of the SMBH disk, the accretion onto AMS still continues and lasts for a timescale of $\tlast=\tviss$.
The duty cycle of the accretion process is $\delta = \tlast/(\tlast+\trej)=4.2\times 10^{-2}$.
The average dimensionless net mass accretion rate of the sMBH is 
$\mathdotM_{\rm ave} = \facc \delta\dotMs=1.1\times 10^3$,
which indicates that mass growing timescale of sMBH is much shorter than the Salpeter time $t_{\rm Salp}=\Ms/\dot{M}_{\rm Edd}=0.45\,\rm Gyr$.
However, the sMBH in the inner region will merge with the SMBH quickly due to the GW radiation \citep[see Equation (24) in ][]{PaperIII}, %from Equation (\ref{eq:tGW}),
which prevents the sMBH mass from growing to be too large.
After the sMBH disk is consumed,
the cavity will quickly cool in a timescale of $\tcoolff$ and the accretion process restarts.

\begin{deluxetable*}{lcccc cc}
\footnotesize
\tablecaption{Explanations and typical values of some key parameters.}
\label{tab:SED}
\tablehead{Parameters & Symbols & Inner & Middle & Outer}
\startdata
\hline
Parameters of the SMBH disk\\
\hline
sMBH locus & $R/(\Rg)$ & $10$ & $10^3$ & $10^5$\\
sMBH mass & $\Ms/(\Msun)$ & $10^3$ & $10^3$ & $10^2$\\
Magnetic field & $B/(\rm G)$ & $5.7\times 10^4$&$2.5\times 10^2$ & $25.2$\\
Opacity & $\kappa/({\rm cm^2\,g^{-1}})$ & $0.4$ & $0.4$ & $1.0$ \\
\hline
Parameters of the sMBH disk\\
\hline
Size & $\Xout/(\rg)$ & $1.2\times 10^3$ & $1.2\times 10^5$ & $7.9\times 10^6$ \\
Viscosity timescale & $\tviss/(\rm days)$ & $6.3\times 10^{-2}$ & $62.6$ & $3.2\times 10^3$ \\
\hline
Parameters of the outflows\\
\hline
Luminosity of the IC scattering & $L_{\rm IC}/(\rm erg\,s^{-1})$ & $4.9\times 10^{42}$ & $2.2\times 10^{43}$ & $6.3\times 10^{41}$\\
Expansion timescale & $\tout/(\rm days)$ & $1.6\times 10^{-2}$ & $1.5$ & $6.3\times 10^3$ &\\
Lasting time & $\tlast/(\rm days)$ & $6.3\times 10^{-2}$ & $62.6$ & $6.3\times 10^3$ &\\
Radius & $\Rout/({\rm cm})$ & $1.2\times 10^{12}$ & $3.6\times 10^{13}$&$6.2\times 10^{15}$ \\
\hline
Parameters of the nonthermal electrons\\
\hline
Number density & $N_0/(\rm cm^{-3})$ & $2.5\times 10^8$ & $2.6\times 10^8$ & $13.4$\\
The maximum of the Lorentz factor & $\gammax$ & $3.3\times 10^3$ & $1.6\times 10^4$ & $2.0\times 10^3$\\
The minimum of the Lorentz factor & $\gammin$ & 1 & 1 & 1\\
The power-law index & $p_{\rm e}$ & 2 & 2 & 2\\
\hline
Parameters of the thermal light curves\\
\hline
Ejecta mass & $M_{\rm ej}/(\Msun)$ & $8.0\times 10^{-5}$ & $0.34$ & $2.4\times 10^2$\\
Diffusion timescale & $\tdiff/(\rm days)$ & $1.4$ & $2.1\times 10^2$ & $6.0\times 10^2$ &\\
Effective light-curve timescale & $\teff/(\rm days)$ & $0.21$ & $24.9$ & $2.8\times 10^3$ &\\
Peak luminosity & $\Lpeak/(\rm erg\,s^{-1})$ & $6.0\times 10^{42}$ & $2.2\times 10^{44}$ & $6.3\times 10^{42}$ &\\
Effective temperature & $\Teff/({\rm K})$ & $2.7\times 10^5$ & $1.2\times 10^5$ & $2.0\times 10^3$\\
\hline
\enddata
\end{deluxetable*}

\section{Observation signatures}
\label{sect:observation}

\subsection{Thermal emission}

Now, we calculate the thermal light curves of the Bondi explosion.
The diffusion timescale of the photons is given by $\tdiff=\kappa M_{\rm ej}/\beta_0 c\Rcav$,
where $M_{\rm ej}=\Mcav$ is the ejecta mass,
and a value of $\beta_0=13.7$ is utilized to accommodate various density profiles of the diffusion mass \citep{Arnett1982}.
For AMS in the outer region, we set $\kappa=1.0$ for simplicity, ignoring its complex dependence on the radius, as shown in Figure \ref{fig:SGdisk}.
For AMSs in the middle and inner regions, the opacity is determined by electron scattering and $\kappa=0.4$.
Since the post-shock medium is quite hot ($\Tsh\sim 10^5, 10^8, 10^9\,$K in the outer, middle, and inner regions, respectively), free-free absorption, bound-free and bound-bound absorption could not be important in the main phase of the Bondi explosion. We may include these effects for the later phase of the explosion.
The cavity expands with a timescale of $\tout$, which is much shorter than the diffusion timescale, as indicated in Table \ref{tab:SED}.
Therefore, an effective light-curve timescale $\teff=\sqrt{2\tdiff \tout}$ is introduced \citep{Chatzopoulos2012}.
The electromagnetic emissions from the Bondi explosion will emerge within the timescale of $\teff$.
We calculate the thermal light curves for a simple form of the input luminosity $L_{\rm inp}=\Lout \theta(\tlast-t)$,
where $\theta(\tlast-t)$ denotes the Heaviside step function.
More detailed interaction processes between the outflows and SMBH disk gas, such as the reverse and forward shock, are beyond the scope of this study.
The output luminosity of the homologously expanding photosphere is described by Equation (3) in \cite{Chatzopoulos2012},
\begin{equation}
\begin{array}{lll}
L_{\rm bol}(t) & = & \displaystyle \frac{2}{\teff}{\rm exp}{\left(-\frac{t^2}{\teff^2}-\frac{2\Rout t}{\vout \teff^2}\right)}
\int^t_0 \left(\frac{\Rout}{\vout \teff}+\frac{t^\prime}{\teff}\right)
\displaystyle 
{\rm exp}\left(\frac{t^{\prime 2}}{\teff^2}+\frac{2\Rout t^\prime}{\vout \teff^2}\right)
L_{\rm inp}(t^\prime)
dt^\prime.
\end{array}
\end{equation}
The initial radius is set as the outflow radius $\Rout$,
and the initial internal energy can be disregarded.

We then obtain the light curves of the thermal emission,
\begin{equation}
\label{eq:lc}
L_{\rm bol}(t) =\left\{ 
\begin{array}{lllc}
\displaystyle
\Lout\left[1-{\rm exp}{\displaystyle \left(-\frac{t^2}{\teff^2}-\frac{2\Rout t}{\vout \teff^2} \right)}\right], & t<\tlast,\\[2.8ex]
\Lout \left[{\rm exp}{\displaystyle \left(\frac{\tlast^2}{\teff^2}+\frac{2\Rout \tlast}{\vout \teff^2}\right)}-1\right]
% &\\
{\rm exp}{\left(\displaystyle -\frac{t^2}{\teff^2}-\frac{2\Rout t}{\vout \teff^2}\right)},
& t>\tlast,
\end{array}
\right. 
\end{equation}
which exhibit an increase over a timescale of $\tlast$
and a Gaussian decay over the effective light-curve timescale $\teff$.
The luminosity reaches its peak after a time interval of $\tlast$ as
$\Lpeak=\Lout[1-{\rm exp}{(-\tlast^2/\teff^2-2\Rout \tlast/\vout \teff^2)}]$,
the detailed calculated results of which are listed in Table \ref{tab:SED}.

Due to the highly effective free-free cooling of the hot gas within the cavity, the expanding shell loses its energy in the form of blackbody radiation with an effective temperature
$\Teff=(\Lpeak/4\pi \Rcav^2 \sigSB)^{1/4}$,
where $\sigSB$ is Stefan-Boltzmann constant.
The blackbody spectra are plotted in purple (see Figures \ref{fig:SEDout}-\ref{fig:SEDinner}).
The thermal luminosity $\Lpeak\approx \therlumOut$ of Bondi explosion in the outer region peaks in the infrared band, slightly higher than that of the AGN disk.
In the middle region, the thermal luminosity of $\Lpeak\approx\therlumMid$ of the Bondi explosion greatly exceeds that of the SMBH disk in the UV band,
while the Bondi explosions occurring in the inner region could lead to soft X-ray flares with a luminosity of $\Lpeak\approx\therlumInn$.

\subsection{Nonthermal Emissions}

As discussed in \S\,\ref{sect:application}, the velocity of the outflows exceeds the local sound speed, leading to the development of shocks \citep{Blandford1987} and the generation of nonthermal electrons and emissions \citep{Amano2022}.
Following the approach of \cite{Inoue1996} and utilizing a simplified homogeneous one-zone model framework,
we compute the broadband SED of synchrotron radiation and IC scattering.
It is worth noting that, unlike the jet case, there is no beaming effect in the case of outflows, and thus, the beaming factor is set to unity.

The thermal electrons are accelerated to a relativistic state due to shock acceleration \citep[see, e.g., ][]{Drury1983, Blandford1987}.
The acceleration timescale is given by $t_{\rm acc}=20 \xi_{\rm acc}R_{\rm L}c/3\vout^2$ \citep{Inoue1996},
where $R_{\rm L}=\gamma \me c^2/eB$ denotes the Larmor radius,
$\me$ is the electron mass,
and $\xi_{\rm acc}$ is a factor characterizing the acceleration efficiency that depends on the acceleration environment.
For instance, $\xi_{\rm acc} \sim 1$ (referred to as the {\it Borm limit}) in supernova remnant \citep{Uchiyama2007},
while in blazars, $\xi_{\rm acc}$ can reach up to $10^7$ \citep{Inoue1996}.
In the context of AMS, where the shock is not relativistic, we take $\xi_{\rm acc} \sim 1$.
The magnetic field is derived as $B=(32\pi a/3)^{1/2} \Tc^2$,
under the assumption of equipartition with the radiation energy density \citep{Burbidge1956},
where $a=4\sigT/c$ is the blackbody radiation constant.
Note that the midplane temperature $\Tc$ in the outer region of AGN disk is approximately a constant $10^4\,\rm K$ (see Figure \ref{fig:SGdisk}).
The nonthermal electrons lose their energy, due to IC scattering in a cooling timescale of $t_{\rm IC}=3\me c/4\sigT\gamma u_{\rm ph}$,
where $\gamma$ is the Lorentz factor of the nonthermal electrons,
and $u_{\rm ph}=a\Tc^4$ is the energy density of the local seed photons radiated by the SMBH disk.
By utilizing the condition $t_{\rm acc}=t_{\rm IC}$, we derive the maximum Lorentz factor $\gammax$.
The energy spectrum of the nonthermal electrons is given by $dN/d\gamma=(1-p_{\rm e})N_0\gamma^{-p_{\rm e}}/(\gammax^{1-p_{\rm e}}-\gammin^{1-p_{\rm e}})$,
where $N_0$ is the total number density obtained by integrating $dN/d\gamma$ from $\gammin$ to $\gammax$.
Here, we set $\gammin=1$ and the spectral index $p_{\rm e}=2$.

Now, we calculate the SED of the nonthermal radiation.
First, we consider the IC scattering,
where the photons from the SMBH disk are scattered by the nonthermal electrons and transformed into $\gamma$-rays.
By setting the IC scattering luminosity as $L_{\rm IC}=\zeta \Lout$,
we can derive the number density of the nonthermal electrons $N_0$.
Here, the parameter $\zeta$ representing the typical fraction of nonthermal emissions generated by shocks is quite uncertain \citep[e.g., ][]{Blandford1987}, so we set $\zeta\sim 0.1$.
The number density of the seed photons is estimated as $n_{\nu}=4\pi B_\nu/h\nu c$,
where $B_\nu = 2h\nu^3/c^2({\rm exp}(h\nu/\kB \Tc)-1)$ is the blackbody spectrum.
Subsequently, we calculate the SEDs of the synchrotron radiation with the magnetic field $B$ and number density of nonthermal electrons $N_0$.
The calculated SEDs of the IC scattering and synchrotron radiation are plotted in Figures \ref{fig:SEDout}-\ref{fig:SEDinner}, with parameters listed in Table \ref{tab:SED}.

\section{Discussion}
\label{sect:discussion}

AMSs in the inner (middle) regions display quasi-periodic eruptions (QPEs) with a duration of several hours (months) and a period of several days (years), providing valuable insights into the QPE with various periods \citep[e.g.,][]{Evans2023, Guolo2024}.
In fact, the AMS model has been applied to Sgr A$^*$ \citep{PaperIII}, successfully explaining its quasi-periodic flickerings observed in the near-infrared band.

The light curves of thermal luminosity for Bondi explosions occurring in different regions of SMBH disk exhibit a broad range of duration timescales, spanning from several hours to decades at different bands from infrared to soft X-ray band.
In the outer region, the diffusion timescale of Bondi explosion is much shorter than the expansion timescale, causing the light curve to behave like the Heaviside step function,
as shown in Figure \ref{fig:SEDout}.
The thermal luminosity can be estimated as $L_{\rm bol}(t)\approx\Lout$ and lasts for a time interval of decades.
The thermal luminosity of Bondi explosions occurring in the middle region exhibits a slow rise and rapid decay (see Figure \ref{fig:SEDmid}).
Conversely, in the inner region, the thermal luminosity of the Bondi explosion experiences a rapid rise and slow decay (see Figure \ref{fig:SEDinner}).
These diverse and intriguing features show that AMSs could be accountable for astronomical transients of varying durations.
Indeed, various atypical transients have been documented in recent years.
For example, \cite{Ofek2021} reported an optical transient, AT 2018lqh, with a duration on the scale of days
attributed to an explosion of low-mass ejecta ($\approx 0.07\,\Msun$),
which is comparable to those of the AMS in the middle region if the sMBH mass and its location are appropriately adjusted.
While in the X-ray band, \cite{Khamitov2023} presented an SRG/eROSITA X-ray catalog with significant proper motions explained by the presence of transient events,
supporting the idea of AMSs being the possible physical origin of these transient events.

The SEDs of Bondi explosions occurring in different regions of the SMBH disk exhibit various features.
The synchrotron radiations of AMS in the outer region could span from $\sim 10\,$GHz to optical bands.
The ratio of radio to optical luminosity is $\sim 10^{-5}$,
which could explain part of the radio emissions of radio-quiet AGNs.
Recent radio sky surveys, such as the Very Large Array Sky Survey \citep[][]{Gordon2021} at 2-4 GHz,
and the LOw-Frequency ARray Two-metre Sky Survey \citep[][]{Shimwell2022} at 120-168 MHz,
can be helpful in the search for radio transients from Bondi explosion.
These catalogs are compared with older sky surveys, such as the NRAO VLA Sky Survey \citep[][]{Condon1998}, the Sydney University Molonglo Sky Survey \citep[][]{Mauch2003}, and Faint Images of the Radio Sky at Twenty Centimeters \citep[][]{Helfand2015}, to study the long timescale radio variability \citep{Nyland2020}. 
In fact, long-term radio variability \citep[e.g., ][]{Hovatta2008, Park2017, Zhang2022} and some radio transients from, such as the Caltech-NRAO Stripe 82 Survey \citep[][]{Mooley2016} with variability timescales between 1 week and 1.5 yr,
the Variables and Slow Transients Survey on the Australian Square Kilometer Array Pathfinder \citep[][]{Murphy2021} with variability timescales from 5 s to 5 yr,
have been reported for a considerable number of AGNs.
These radio transient events are very helpful to search for the AMSs.

Another interesting feature of the AMS SED is the high-energy $\gamma$-ray photons (peaked from approximately $10$ MeV to GeV).
For AMSs located in the middle regions of the SMBH disk, the $\gamma$-ray luminosity is estimated to be around $10^{43}\,\rm erg\, s^{-1}$,
corresponding to a flux of $3.7\times 10^{-13}\,\rm erg\,s^{-1}\,cm^{-2}$ for a nearby AGN with redshift $z=0.1$,
which is comparable with the sensitivity of the Fermi/LAT at the GeV band with 10 years of 
observation\footnote{\url{https://www.slac.stanford.edu/exp/glast/groups/canda/lat_Performance.htm}}.
It is worth noting that the $\gamma$-ray luminosity is determined by several physical parameters, for example, as described in Equations (\ref{eq:XoutMid}) and (\ref{eq:LoutMid}),
a higher sMBH mass results in a shorter size of the sMBH disk and more powerful outflows,
leading to higher $\gamma$-ray luminosity and flux.
On the other hand, the stacking technique of $\gamma$-rays presented in \cite{Paliya2019} and \cite{Ajello2021} will be useful for detecting fainter $\gamma$-rays below the Fermi/LAT sensitivity.
In fact, significant $\gamma$-ray detections have been reported for some nearby low-luminosity active galactic nuclei \citep[LLAGN; ][]{Menezes2020}.
% where the luminosities of NGC 4486 and NGC 4261, for example, are approximately $10^{42}\,\rm erg\,s^{-1}$.
It is easy for the AMSs to outshine from the LLAGN,
although SMBHs in LLAGN generally accrete with low, sub-Eddington accretion rates \citep{Ho2008},
which influence the AMS luminosity (see Equations (\ref{eq:LoutOut}), (\ref{eq:LoutMid}), and (\ref{eq:LoutInn})).

\section{Conclusion}
\label{sect:conclusion}
In this study, we investigate the AMSs embedded in different regions of the disk surrounding the SMBHs, 
namely, the inner region (typical radius $R=10\,\Rg$), the middle region (typical radius $R=10^3\,\Rg$), and the outer region(typical radius $R=10^5\,\Rg$).
In the inner and middle regions, the Toomre parameter, $Q>1$, and the standard model \citep{Shakura1973} is utilized.
In the outer region, $Q\approx1$, the self-gravitating disk model \citep{Goodman2003, Sirko2003} is employed.
Bondi explosions in these regions exhibit both similarities and distinct characteristics,
depending on various gas environments, primarily including gas density, sound speed, and half-thickness of the SMBH disk.
The main findings are summarized below:

(1) The main physical processes are similar for AMSs embedded in the inner and middle regions of the SMBH disk.
The AMS experiences hyper-Eddington accretion ($\sim 10^6-10^7\,L_{\rm Edd, s}/c^2$),
resulting in the development of strong outflows that collide with the SMBH disk, generating shocks, heating the SMBH disk gas, accelerating electrons to relativistic state, and emitting nonthermal radiation.
In the inner and middle regions, after the powerful outflow rushes out of the SMBH disk, a cavity is formed, but it quickly cools and is then refilled by the surrounding cool gas of the SMBH disk.
While in the outer region, the inflows of Bondi accretion and outflows from Bondi explosion coexist and can last for several decades, as shown in Figure \ref{fig:cartoon}.

(2) We compute the thermal light curves for a constant input luminosity within the lasting timescale of the outflows.
The results show that the flare of the Bondi explosion in the inner region displays fast rise and slow Gaussian decay, with a timescale of several hours and a luminosity of $\sim \therlumInn$ peaked at the soft X-ray band.
While in the middle region, the Bondi explosion exhibits slow rise and rapid Gaussian decay, with a timescale of months and a luminosity of $\sim \therlumMid$ peaked at the UV band.
In the outer region, the light curve of Bondi explosion resembles the Heaviside step function, lasting for decades and contributing a slightly higher luminosity of $\sim \therlumOut$ than the AGN disk itself in the infrared band.
These light curves provide valuable insights into the diverse astronomical transient events associated with AGNs.

(3) We calculate the multiwavelength SEDs of the Bondi explosion from radio to $\gamma$-ray bands.
The $\gamma$-rays luminosity of IC ranges from $\gamlumOut-\gamlumMid$ in the GeV (middle and inner regions) and $10$ MeV (outer region) bands, respectively.
Moreover, the radio emission due to synchrotron radiation from Bondi explosion occurring in the outer region can contribute to that of the radio-quiet AGNs or result in long-term radio variability over a timescale of several decades.
The synchrotron radiation of Bondi explosion in the middle and inner regions peaks at the X-ray band with luminosities of $\sim \synlumMid$ and $\sim \synlumInn$, respectively.

\begin{acknowledgments}
%The authors thank an anonymous referee for a helpful report.
IHEP AGN Group members are acknowledged for useful discussions.
Many thanks to Fu-Lin Li for useful discussions about the supernova light-curve calculations.
We acknowledge financial support from the National Key R\&D Program of China (2021YFA1600404), the National Natural Science Foundation of China (NSFC-11991050, -11991054,-12333003).
\end{acknowledgments}

\appendix

\begin{figure*}
\centering
\includegraphics[scale=0.85, trim=50 0 0 0]{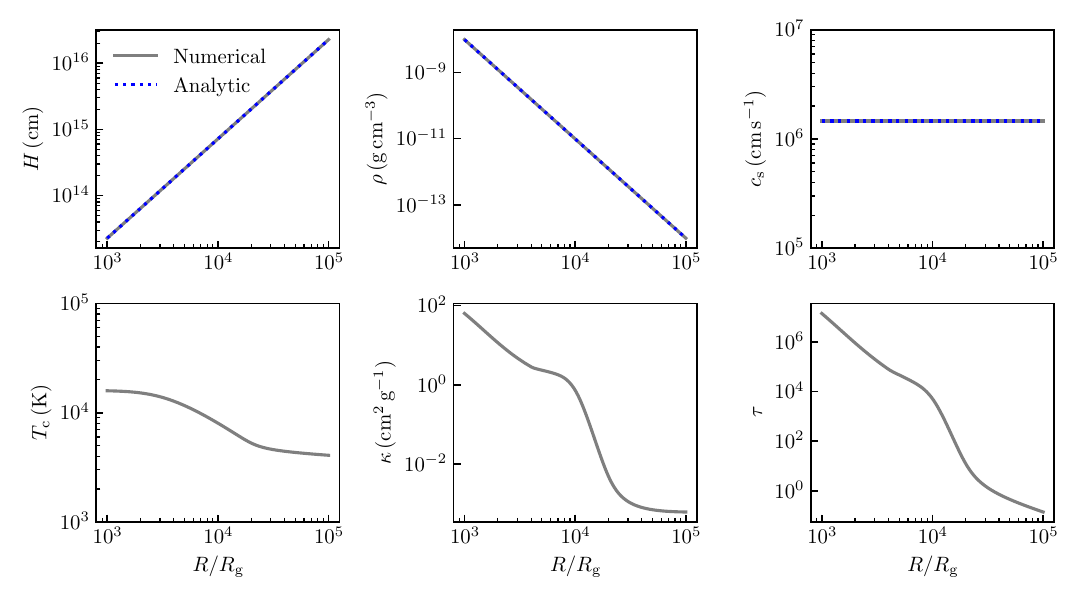}
\caption{\footnotesize
The parameters of the self-gravitating disk for given $M_8=1$, $\dotMp$=1, $\alpha=0.1$, and $b=0$.
The blue dotted lines denote the analytic solutions described by Equations (\ref{eq:HSG})-(\ref{eq:csSG}).
The gray lines denote the numerical solutions derived from Equations (\ref{eq:rhoSGeq})-(\ref{eq:vrpeq}).}
\label{fig:SGdisk}
\end{figure*}

\section{Hyper-Eddington Accretion Disk}
\label{appdx: sMBHdisk}

In the region close to the sMBH, the falling gas will form a hyper-Eddington accretion disk \citep[e.g.,][]{PaperI, ChenKen2023a}
since the differential rotation of gas leads to net AM.
The outer radius of sMBH disk $\Xout$ can be estimated through the conservation of AM.
The specific net AM of the gas from differential rotation is given by
$\Delta \ell_{\rm p}=(\Delta R)^2\OmgK/2$,
where $\Delta R$ is the size of the accretion region.
The specific AM of the sMBH disk is $\Delta \ell_{\rm s}=\sqrt{\Xout G\Ms}$.
By setting $\Delta \ell_{\rm p}=\Delta \ell_{\rm s}$,
we derive 
\begin{equation}
\label{eq:Xout}
\Xout=\frac{(\Delta R)^4\Mp}{4 R^3\Ms}.
\end{equation}
The sMBH disk size strongly depends on $\Delta R$.

A radius-dependent accretion rate is often adopted in references,
$\dot{M} = \dot{M}_{\rm out} \left(r/\Xout\right)^{s}$,
where $\dot{M}_{\rm out}=\dotMs\dot{M}_{\rm Edd}$ is the accretion rate at outer radius $\Xout$,
% estimated through the Equation (\ref{eq:dotMs}),
and $r$ is the radius of the sMBH disk.
The index $s$ ranges from 0 to 1,
which is dependent on the ratio of the radiation energy to the dissipated gravitational energy \citep{Begelman2012}.
In purely adiabatic flows, $s=1$ \citep[][]{Begelman2012}.
$s$ can also be estimated by numerical results in different cases \citep[e.g.,][]{Yang2014,Kitaki2021,Pan2021, Cao2022}.
% The self-similar solutions of disk with outflow are derived by \cite{Blandford2004}.
For example, \cite{Yang2014} considered the effect the $T_{\theta \phi}$ component of viscous stress in a two-dimensional simulation and found that $s$ is in the range of $\sim 0.4-1.0$.
In consideration of the above uncertainties, a moderate index $s=1/2$ is adopted in this work.
So we have the fraction of the accretion rate onto the sMBH to the Bondi accretion rate,
\begin{equation}
\label{eq:facc}
\facc =\left(\frac{X_{\rm in}}{\Xout}\right)^{1/2},
\end{equation}
where $X_{\rm in}=6\,\rg$ is the inner radius of the sMBH disk.

Based on equations (1)-(7) in \cite{Wang1999} but replacing $\dot{M}$ with $\dot{M}_{\rm out}(r/\Xout)^s$,
the governing equations read as
\begin{equation}
\label{eq:slim1}
\dot{M}_{\rm out}\left(\frac{r}{\Xout}\right)^{s}
=4\pi r\Hs\rhos \vrs,
\end{equation}
\begin{equation}
\frac{\Ps}{\rhos}=\Hs^2\OmgKs^2,
\end{equation}
\begin{equation}
\dot{M}_{\rm out}\left(\frac{r}{\Xout}\right)^{s}
\left(l-l_{\rm in}\right)
=4\pi r^2 \Hs\alpha \Ps,
\end{equation}
\begin{equation}
\frac{1}{\rhos}\frac{d\Ps}{dr}-\left(\Omega^2-\OmgKs^2\right)r
+\vrs\frac{d\vrs}{dr}+\frac{\Ps}{\rhos}\frac{d\ln \OmgKs}{dr}=0,
\end{equation}
\begin{equation}
Q_{\rm adv}=\frac{\dot{M}_{\rm out}}{4\pi r^2}\left(\frac{r}{\Xout}\right)^{s}\left(\frac{\Ps}{\rhos}\right)\xi,
\end{equation}
\begin{equation}
Q_{\rm vis}=\frac{nf\Omega^2\dot{M}_{\rm out}}{4\pi}\left(\frac{r}{\Xout}\right)^{s},
\end{equation}
\begin{equation}
Q_{\rm adv}=Q_{\rm vis},
\end{equation}
\begin{equation}
\label{eq:slim8}
\Ps=\frac{a\Ts^4}{3},
\end{equation}
where $l=\Omega r^2$ is the specific AM,
$l_{\rm in}$ is the specific AM at the inner boundary,
$\OmgKs=\sqrt{G\Ms/r^3}$ is the Keplerian angular velocity of the sMBH disk,
$\xi=(4-0.75\beta)\gamma_{\rho}-(12-10.5\beta)\gamma_{\rm T}$,
$\beta \ll 1$ is the ratio of gas to total pressure,
$\gamma_{\rho}=d\ln \rhos/d\ln r$,
$\gamma_{\rm T}=d\ln \Ts/d\ln r$,
$n=-d\ln \Omega/d\ln r$,
$f=1-l_{\rm in}/l$,
and $a$ is the blackbody radiation constant.
Note that the coefficients $B_i$ in \cite{Muchotrzeb1982} are around the unit and can be disregarded following \cite{Wang1999}.
There are 8 equations of the sMBH disk to determine 8 free physical quantities, i.e.,
total pressure $\Ps$,
density $\rhos$,
radial velocity $\vrs$,
disk half-thickness $\Hs$,
angular velocity $\Omega$,
advection cooling rate $Q_{\rm adv}$,
energy generation per area $Q_{\rm vis}$,
and temperature $\Ts$.

The boundary condition of zero-torque at the inner edge of the disk is adopted here.
By combining Equations (\ref{eq:slim1})-(\ref{eq:slim8}), we then derive the self-similar solutions of hyper-Eddington disk with outflows,
\begin{equation}
\Ps=\left(\frac{\xi f}{n}\right)^{1/2}\frac{\OmgKs \dot{M}_{\rm out}}{4\pi\alpha r}\left(\frac{r}{\Xout}\right)^{s}
\propto r^{s-5/2},
\end{equation}
\begin{equation}
\rhos=\displaystyle\frac{\gamma_0^2}{4\pi\alpha}
\left(\frac{\xi^3}{n^3f}\right)^{1/2}
\frac{\dot{M}_{\rm out}}{\OmgKs r^3}
\left(\frac{r}{\Xout}\right)^s
\propto r^{s-3/2},
\end{equation}
\begin{equation}
\vrs=\displaystyle\frac{n\alpha}{\xi \gamma_0}r\OmgKs
\propto r^{-1/2},
\end{equation}
\begin{equation}
\Hs=\frac{r}{\gamma_0}\left(\frac{nf}{\xi}\right)^{1/2}
\propto r,
\end{equation}
\begin{equation}
\Omega=\displaystyle\frac{\OmgKs}{\gamma_0}
\propto r^{-3/2},
\end{equation}
\begin{equation}
\Ts=\displaystyle\left(\frac{\xi f}{n}\right)^{1/8}
\left(\frac{3\dot{M}_{\rm out}\OmgKs}{4\pi a \alpha r}\right)^{1/4}
\left(\frac{r}{\Xout}\right)^{s/4}
\propto r^{s/4-5/8},
\end{equation}
where $\gamma_{\rho}=d\ln \rhos/d\ln r=s-3/2$,
$\gamma_{\rm T}=d\ln \Ts/d\ln r=s/4-5/8$,
$n=-d\ln \Omega/d\ln r=3/2$,
$\gamma_0=[1-nf(\gamma_p-3/2)/\xi-n^2\alpha^2\gamma_v/\xi^2]$,
$\gamma_p=d\ln \Ps/d\ln r=s-5/2$,
$\gamma_v=d\ln \vrs/d\ln r=-1/2$.
By setting $r=\Xout$, the radial velocity $\vrs$ and viscosity timescale of the sMBH disk can be estimated as
\begin{equation}
\label{eq:vrs}
\vrs\approx \frac{1}{25}\sqrt{\frac{G\Ms}{\Xout}},
\end{equation}
\begin{equation}
\label{eq:tviss}
\tviss = \frac{\Xout}{\vrs}.
\end{equation}

\setcounter{equation}{0}
\section{self-gravitating accretion disk}
\label{appdx: SGdisk}

In the outer part of the AGN disk, the accretion disk becomes self-gravity dominated \citep{Toomre1964}.
We consider the regions where the Toomre parameter $Q\sim 1$.
The basic equations read \citep{Goodman2003, Sirko2003} as
\begin{equation}\label{eq:rhoSGeq}
\rhop=\displaystyle \frac{\Omega^2}{2\pi GQ},
\end{equation}
\begin{equation}\label{eq:csSGeq}
\beta^b \cs^2\Sigp=\displaystyle \frac{\dot{M}_{\rm p}\Omega}{3\pi \alpha},
\end{equation} 
\begin{equation}\label{eq:SigmaSGeq}
\Sigp=\displaystyle 2\rhop \Hp,
\end{equation} 
\begin{equation}\label{eq:HSGeq}
\Hp=\displaystyle \frac{\cs}{\Omega},
\end{equation} 
\begin{equation}\label{eq:Tceq}
\Tc^4=\displaystyle \left(\frac{3\tau}{8}+\frac{1}{2}+\frac{1}{4\tau}\right)\Teff^4,\\
\end{equation} 
\begin{equation}
\tau= \displaystyle \frac{\kappa \Sigp}{2},
\end{equation} 
\begin{equation}
p_{\rm rad}=\displaystyle \frac{\tau \sigSB \Teff^4}{2c},
\end{equation} 
\begin{equation}
p_{\rm gas}=\displaystyle \frac{\rhop k_{\rm B}\Tc}{m},
\end{equation} 
\begin{equation}
\beta=\displaystyle \frac{p_{\rm gas}}{p_{\rm gas}+p_{\rm rad}},
\end{equation} 
\begin{equation}
\cs^2=\displaystyle \frac{p_{\rm gas}+p_{\rm rad}}{\rhop},
\end{equation}
\begin{equation}\label{eq:kappaeq}
\kappa = \kappa(\rhop,\,\Tc),
\end{equation}
\begin{equation}
\label{eq:vrpeq}
\dot{M}_{\rm p}=-2\pi R \vrp\Sigp,
\end{equation}
where $\Omega=\sqrt{G\Mp/R^3}$ is the Keplerian angular velocity,
and $\sigSB$ is the Stefan-Boltzmann constant,
and $m=0.62\,m_{\rm H}$ is the mean molecular mass.
We add the mass conservation equation (\ref{eq:vrpeq}) compared with \cite{Sirko2003} to compute the radial velocity $\vrp$.
The parameter $b=0 \,(b=1)$ means that the viscosity is proportional to the total (gas) pressure.
For a given mass accretion rate $\dot{M}_{\rm p}$, the viscosity parameter $\alpha$, SMBH mass $\Mp$, radius $R$, minimum Toomre parameter $Q_{\rm min}$, and $b$, these 12 equations can be solved to derive the 12 unknown parameters,
gas density $\rhop$, midplane temperature $\Tc$, optical depth $\tau$,
effective temperature $\Teff$, opacity $\kappa$, surface density $\Sigp$,
fraction of gas pressure $\beta$, sound speed $\cs$, radiation pressure $p_{\rm rad}$,
gas pressure $p_{\rm gas}$, disk half-thickness $\Hp$, and radial velocity $\vrp$.

In this paper, $b=0$ is considered, i.e., the viscosity proportional to the total pressure.
For given $\dot{M}_{\rm p}$, $\alpha$, $\Mp$, $R$, and $Q_{\rm min}=1$, the solutions of $\Hp$, $\rhop$, $\cs$, and $\Sigp$ can be derived by combining Equations (\ref{eq:rhoSGeq})-(\ref{eq:HSGeq}),
\begin{equation}
\label{eq:HSG}
\Hp=\displaystyle \frac{\dot{M}_{\rm p}^{1/3}R^{3/2}}{(3\alpha)^{1/3}G^{1/6}\Mp^{1/2}},
\end{equation}

\begin{equation}
\label{eq:rhoSG}
\rhop=\displaystyle\frac{\Mp}{2\pi R^3},
\end{equation}

\begin{equation}
\label{eq:csSG}
\cs=\displaystyle \left(\frac{G\dot{M}_{\rm p}}{3\alpha}\right)^{1/3},
\end{equation}

\begin{equation}
\label{eq:SigmaSG}
\Sigp=\displaystyle \frac{\Mp^{1/2}\dot{M}_{\rm p}^{1/3}}{\pi(3\alpha)^{1/3} G^{1/6}R^{3/2}}.
\end{equation}
\vspace{0em}

Based on Equation (\ref{eq:vrpeq}) and the solution of surface density (Equation (\ref{eq:SigmaSG})), the radial velocity $\vrp$ is derived as
\begin{equation}
\label{eq:vpSG}
\vrp=\frac{(3\alpha)^{1/3}G^{1/6}\dot{M}_{\rm p}^{2/3}R^{1/2}}{2\Mp^{1/2}}.
\end{equation}

The remaining 7 parameters, $\Tc$, $\tau$, $\Teff$, $\kappa$, $\beta$, $p_{\rm rad}$, and $p_{\rm gas}$,
can be obtained using the above solutions and Equations (\ref{eq:Tceq})-(\ref{eq:kappaeq}). The key physical quantity, opacity $\kappa$ determined by the temperature and density of the disk gas,
can not be expressed analytically \citep{Alexander1994}.
So, we derive their numerical solutions and plot them in Figure \ref{fig:SGdisk}.

\bibliography{ref}{}

\begin{thebibliography}{}
\makeatletter
\relax
\def\mn@urlcharsother{\let\do\@makeother \do\$\do\&\do\#\do\^\do\_\do\%\do\~}
\def\mn@doi{\begingroup\mn@urlcharsother \@ifnextchar [ {\mn@doi@}
  {\mn@doi@[]}}
\def\mn@doi@[#1]#2{\def\@tempa{#1}\ifx\@tempa\@empty \href
  {http://dx.doi.org/#2} {doi:#2}\else \href {http://dx.doi.org/#2} {#1}\fi
  \endgroup}
\def\mn@eprint#1#2{\mn@eprint@#1:#2::\@nil}
\def\mn@eprint@arXiv#1{\href {http://arxiv.org/abs/#1} {{\tt arXiv:#1}}}
\def\mn@eprint@dblp#1{\href {http://dblp.uni-trier.de/rec/bibtex/#1.xml}
  {dblp:#1}}
\def\mn@eprint@#1:#2:#3:#4\@nil{\def\@tempa {#1}\def\@tempb {#2}\def\@tempc
  {#3}\ifx \@tempc \@empty \let \@tempc \@tempb \let \@tempb \@tempa \fi \ifx
  \@tempb \@empty \def\@tempb {arXiv}\fi \@ifundefined
  {mn@eprint@\@tempb}{\@tempb:\@tempc}{\expandafter \expandafter \csname
  mn@eprint@\@tempb\endcsname \expandafter{\@tempc}}}

\bibitem[\protect\citeauthoryear{{Abbott} et~al.,}{{Abbott}
  et~al.}{2020}]{Abbott2020}
{Abbott} R.,  et~al., 2020, \mn@doi [\prl] {10.1103/PhysRevLett.125.101102},
  \href {https://ui.adsabs.harvard.edu/abs/2020PhRvL.125j1102A} {125, 101102}

\bibitem[\protect\citeauthoryear{{Ajello} et~al.,}{{Ajello}
  et~al.}{2021}]{Ajello2021}
{Ajello} M.,  et~al., 2021, \mn@doi [\apj] {10.3847/1538-4357/ac1bb2}, \href
  {https://ui.adsabs.harvard.edu/abs/2021ApJ...921..144A} {921, 144}

\bibitem[\protect\citeauthoryear{{Alexander} \& {Ferguson}}{{Alexander} \&
  {Ferguson}}{1994}]{Alexander1994}
{Alexander} D.~R.,  {Ferguson} J.~W.,  1994, \mn@doi [\apj] {10.1086/175039},
  \href {https://ui.adsabs.harvard.edu/abs/1994ApJ...437..879A} {437, 879}

\bibitem[\protect\citeauthoryear{{Ali-Dib} \& {Lin}}{{Ali-Dib} \&
  {Lin}}{2023}]{Ali-Dib2023}
{Ali-Dib} M.,  {Lin} D. N.~C.,  2023, \mn@doi [\mnras]
  {10.1093/mnras/stad2774}, \href
  {https://ui.adsabs.harvard.edu/abs/2023MNRAS.526.5824A} {526, 5824}

\bibitem[\protect\citeauthoryear{{Amano} et~al.,}{{Amano}
  et~al.}{2022}]{Amano2022}
{Amano} T.,  et~al., 2022, \mn@doi [Reviews of Modern Plasma Physics]
  {10.1007/s41614-022-00093-1}, \href
  {https://ui.adsabs.harvard.edu/abs/2022RvMPP...6...29A} {6, 29}

\bibitem[\protect\citeauthoryear{{Arnett}}{{Arnett}}{1982}]{Arnett1982}
{Arnett} W.~D.,  1982, \mn@doi [\apj] {10.1086/159681}, \href
  {https://ui.adsabs.harvard.edu/abs/1982ApJ...253..785A} {253, 785}

\bibitem[\protect\citeauthoryear{{Artymowicz}, {Lin}  \&
  {Wampler}}{{Artymowicz} et~al.}{1993}]{Artymowicz1993}
{Artymowicz} P.,  {Lin} D.~N.~C.,   {Wampler} E.~J.,  1993, \mn@doi [\apj]
  {10.1086/172690}, \href
  {https://ui.adsabs.harvard.edu/abs/1993ApJ...409..592A} {409, 592}

\bibitem[\protect\citeauthoryear{{Ashton}, {Ackley}, {Hernandez}  \&
  {Piotrzkowski}}{{Ashton} et~al.}{2021}]{Ashton2021}
{Ashton} G.,  {Ackley} K.,  {Hernandez} I.~M.,   {Piotrzkowski} B.,  2021,
  \mn@doi [Classical and Quantum Gravity] {10.1088/1361-6382/ac33bb}, \href
  {https://ui.adsabs.harvard.edu/abs/2021CQGra..38w5004A} {38, 235004}

\bibitem[\protect\citeauthoryear{{Bartos}, {Kocsis}, {Haiman}  \&
  {M{\'a}rka}}{{Bartos} et~al.}{2017}]{Bartos2017}
{Bartos} I.,  {Kocsis} B.,  {Haiman} Z.,   {M{\'a}rka} S.,  2017, \mn@doi
  [\apj] {10.3847/1538-4357/835/2/165}, \href
  {https://ui.adsabs.harvard.edu/abs/2017ApJ...835..165B} {835, 165}

\bibitem[\protect\citeauthoryear{{Baruteau}, {Cuadra}  \& {Lin}}{{Baruteau}
  et~al.}{2011}]{Baruteau2011}
{Baruteau} C.,  {Cuadra} J.,   {Lin} D.~N.~C.,  2011, \mn@doi [\apj]
  {10.1088/0004-637X/726/1/28}, \href
  {https://ui.adsabs.harvard.edu/abs/2011ApJ...726...28B} {726, 28}

\bibitem[\protect\citeauthoryear{{Begelman}}{{Begelman}}{2012}]{Begelman2012}
{Begelman} M.~C.,  2012, \mn@doi [\mnras] {10.1111/j.1365-2966.2011.20071.x},
  \href {https://ui.adsabs.harvard.edu/abs/2012MNRAS.420.2912B} {420, 2912}

\bibitem[\protect\citeauthoryear{{Bellovary}, {Mac Low}, {McKernan}  \&
  {Ford}}{{Bellovary} et~al.}{2016}]{Bellovary2016}
{Bellovary} J.~M.,  {Mac Low} M.-M.,  {McKernan} B.,   {Ford} K.~E.~S.,  2016,
  \mn@doi [\apjl] {10.3847/2041-8205/819/2/L17}, \href
  {https://ui.adsabs.harvard.edu/abs/2016ApJ...819L..17B} {819, L17}

\bibitem[\protect\citeauthoryear{{Blandford} \& {Eichler}}{{Blandford} \&
  {Eichler}}{1987}]{Blandford1987}
{Blandford} R.,  {Eichler} D.,  1987, \mn@doi [\physrep]
  {10.1016/0370-1573(87)90134-7}, \href
  {https://ui.adsabs.harvard.edu/abs/1987PhR...154....1B} {154, 1}

\bibitem[\protect\citeauthoryear{{Bondi}}{{Bondi}}{1952}]{Bondi1952}
{Bondi} H.,  1952, \mn@doi [\mnras] {10.1093/mnras/112.2.195}, \href
  {https://ui.adsabs.harvard.edu/abs/1952MNRAS.112..195B} {112, 195}

\bibitem[\protect\citeauthoryear{{Boyce} et~al.,}{{Boyce}
  et~al.}{2022}]{Boyce2022}
{Boyce} H.,  et~al., 2022, \mn@doi [\apj] {10.3847/1538-4357/ac6104}, \href
  {https://ui.adsabs.harvard.edu/abs/2022ApJ...931....7B} {931, 7}

\bibitem[\protect\citeauthoryear{{Burbidge}}{{Burbidge}}{1956}]{Burbidge1956}
{Burbidge} G.~R.,  1956, \mn@doi [\apj] {10.1086/146237}, \href
  {https://ui.adsabs.harvard.edu/abs/1956ApJ...124..416B} {124, 416}

\bibitem[\protect\citeauthoryear{{Burleigh}, {McKee}, {Cunningham}, {Lee}  \&
  {Klein}}{{Burleigh} et~al.}{2017}]{Burleigh2017}
{Burleigh} K.~J.,  {McKee} C.~F.,  {Cunningham} A.~J.,  {Lee} A.~T.,   {Klein}
  R.~I.,  2017, \mn@doi [\mnras] {10.1093/mnras/stx439}, \href
  {https://ui.adsabs.harvard.edu/abs/2017MNRAS.468..717B} {468, 717}

\bibitem[\protect\citeauthoryear{{Cantiello}, {Jermyn}  \& {Lin}}{{Cantiello}
  et~al.}{2021}]{Cantiello2021}
{Cantiello} M.,  {Jermyn} A.~S.,   {Lin} D. N.~C.,  2021, \mn@doi [\apj]
  {10.3847/1538-4357/abdf4f}, \href
  {https://ui.adsabs.harvard.edu/abs/2021ApJ...910...94C} {910, 94}

\bibitem[\protect\citeauthoryear{{Cao} \& {Gu}}{{Cao} \& {Gu}}{2022}]{Cao2022}
{Cao} X.,  {Gu} W.-M.,  2022, \mn@doi [\apj] {10.3847/1538-4357/ac8980}, \href
  {https://ui.adsabs.harvard.edu/abs/2022ApJ...936..141C} {936, 141}

\bibitem[\protect\citeauthoryear{{Chatzopoulos}, {Wheeler}  \&
  {Vinko}}{{Chatzopoulos} et~al.}{2012}]{Chatzopoulos2012}
{Chatzopoulos} E.,  {Wheeler} J.~C.,   {Vinko} J.,  2012, \mn@doi [\apj]
  {10.1088/0004-637X/746/2/121}, \href
  {https://ui.adsabs.harvard.edu/abs/2012ApJ...746..121C} {746, 121}

\bibitem[\protect\citeauthoryear{{Chen} \& {Dai}}{{Chen} \&
  {Dai}}{2024}]{Chen2024}
{Chen} K.,  {Dai} Z.-G.,  2024, \mn@doi [\apj] {10.3847/1538-4357/ad0dfd},
  \href {https://ui.adsabs.harvard.edu/abs/2024ApJ...961..206C} {961, 206}

\bibitem[\protect\citeauthoryear{{Chen} \& {Lin}}{{Chen} \&
  {Lin}}{2023}]{Chen2023}
{Chen} Y.-X.,  {Lin} D. N.~C.,  2023, \mn@doi [\mnras] {10.1093/mnras/stad992},
  \href {https://ui.adsabs.harvard.edu/abs/2023MNRAS.522..319C} {522, 319}

\bibitem[\protect\citeauthoryear{{Chen} \& {Lin}}{{Chen} \&
  {Lin}}{2024}]{ChenLin2024}
{Chen} Y.-X.,  {Lin} D. N.~C.,  2024, \mn@doi [\apj]
  {10.3847/1538-4357/ad3c3a}, \href
  {https://ui.adsabs.harvard.edu/abs/2024ApJ...967...88C} {967, 88}

\bibitem[\protect\citeauthoryear{{Chen}, {Ren}  \& {Dai}}{{Chen}
  et~al.}{2023}]{ChenKen2023a}
{Chen} K.,  {Ren} J.,   {Dai} Z.-G.,  2023, \mn@doi [\apj]
  {10.3847/1538-4357/acc45f}, \href
  {https://ui.adsabs.harvard.edu/abs/2023ApJ...948..136C} {948, 136}

\bibitem[\protect\citeauthoryear{{Cheng} \& {Wang}}{{Cheng} \&
  {Wang}}{1999}]{Cheng1999}
{Cheng} K.~S.,  {Wang} J.-M.,  1999, \mn@doi [\apj] {10.1086/307572}, \href
  {https://ui.adsabs.harvard.edu/abs/1999ApJ...521..502C} {521, 502}

\bibitem[\protect\citeauthoryear{{Condon}, {Cotton}, {Greisen}, {Yin},
  {Perley}, {Taylor}  \& {Broderick}}{{Condon} et~al.}{1998}]{Condon1998}
{Condon} J.~J.,  {Cotton} W.~D.,  {Greisen} E.~W.,  {Yin} Q.~F.,  {Perley}
  R.~A.,  {Taylor} G.~B.,   {Broderick} J.~J.,  1998, \mn@doi [\aj]
  {10.1086/300337}, \href
  {https://ui.adsabs.harvard.edu/abs/1998AJ....115.1693C} {115, 1693}

\bibitem[\protect\citeauthoryear{{Davies} \& {Lin}}{{Davies} \&
  {Lin}}{2020}]{Davies2020}
{Davies} M.~B.,  {Lin} D. N.~C.,  2020, \mn@doi [\mnras]
  {10.1093/mnras/staa2590}, \href
  {https://ui.adsabs.harvard.edu/abs/2020MNRAS.498.3452D} {498, 3452}

\bibitem[\protect\citeauthoryear{{Dittmann}, {Cantiello}  \&
  {Jermyn}}{{Dittmann} et~al.}{2021}]{Dittmann2021}
{Dittmann} A.~J.,  {Cantiello} M.,   {Jermyn} A.~S.,  2021, \mn@doi [\apj]
  {10.3847/1538-4357/ac042c}, \href
  {https://ui.adsabs.harvard.edu/abs/2021ApJ...916...48D} {916, 48}

\bibitem[\protect\citeauthoryear{{Dittmann}, {Dempsey}  \& {Li}}{{Dittmann}
  et~al.}{2024}]{Dittmann2024}
{Dittmann} A.~J.,  {Dempsey} A.~M.,   {Li} H.,  2024, \mn@doi [\apj]
  {10.3847/1538-4357/ad23ce}, \href
  {https://ui.adsabs.harvard.edu/abs/2024ApJ...964...61D} {964, 61}

\bibitem[\protect\citeauthoryear{{Drury}}{{Drury}}{1983}]{Drury1983}
{Drury} L.~O.,  1983, \mn@doi [Reports on Progress in Physics]
  {10.1088/0034-4885/46/8/002}, \href
  {https://ui.adsabs.harvard.edu/abs/1983RPPh...46..973D} {46, 973}

\bibitem[\protect\citeauthoryear{{Evans} et~al.,}{{Evans}
  et~al.}{2023}]{Evans2023}
{Evans} P.~A.,  et~al., 2023, \mn@doi [Nature Astronomy]
  {10.1038/s41550-023-02073-y}, \href
  {https://ui.adsabs.harvard.edu/abs/2023NatAs...7.1368E} {7, 1368}

\bibitem[\protect\citeauthoryear{{Fan} \& {Wu}}{{Fan} \& {Wu}}{2023}]{Fan2023}
{Fan} X.,  {Wu} Q.,  2023, \mn@doi [\apj] {10.3847/1538-4357/acb532}, \href
  {https://ui.adsabs.harvard.edu/abs/2023ApJ...944..159F} {944, 159}

\bibitem[\protect\citeauthoryear{{Frank}, {King}  \& {Raine}}{{Frank}
  et~al.}{2002}]{Frank2002}
{Frank} J.,  {King} A.,   {Raine} D.~J.,  2002, {Accretion Power in
  Astrophysics: Third Edition}.
Cambridge, UK: Cambridge University Press

\bibitem[\protect\citeauthoryear{{GRAVITY Collaboration} et~al.,}{{GRAVITY
  Collaboration} et~al.}{2020}]{GRAVITY2020}
{GRAVITY Collaboration} et~al., 2020, \mn@doi [\aap]
  {10.1051/0004-6361/201937233}, \href
  {https://ui.adsabs.harvard.edu/abs/2020A&A...635A.143G} {635, A143}

\bibitem[\protect\citeauthoryear{{Genzel}, {Sch{\"o}del}, {Ott}, {Eckart},
  {Alexander}, {Lacombe}, {Rouan}  \& {Aschenbach}}{{Genzel}
  et~al.}{2003}]{Genzel2003}
{Genzel} R.,  {Sch{\"o}del} R.,  {Ott} T.,  {Eckart} A.,  {Alexander} T.,
  {Lacombe} F.,  {Rouan} D.,   {Aschenbach} B.,  2003, \mn@doi [\nat]
  {10.1038/nature02065}, \href
  {https://ui.adsabs.harvard.edu/abs/2003Natur.425..934G} {425, 934}

\bibitem[\protect\citeauthoryear{{Genzel}, {Eisenhauer}  \&
  {Gillessen}}{{Genzel} et~al.}{2010}]{Genzel2010}
{Genzel} R.,  {Eisenhauer} F.,   {Gillessen} S.,  2010, \mn@doi [Reviews of
  Modern Physics] {10.1103/RevModPhys.82.3121}, \href
  {https://ui.adsabs.harvard.edu/abs/2010RvMP...82.3121G} {82, 3121}

\bibitem[\protect\citeauthoryear{{Goodman}}{{Goodman}}{2003}]{Goodman2003}
{Goodman} J.,  2003, \mn@doi [\mnras] {10.1046/j.1365-8711.2003.06241.x}, \href
  {https://ui.adsabs.harvard.edu/abs/2003MNRAS.339..937G} {339, 937}

\bibitem[\protect\citeauthoryear{{Gordon} et~al.,}{{Gordon}
  et~al.}{2021}]{Gordon2021}
{Gordon} Y.~A.,  et~al., 2021, \mn@doi [\apjs] {10.3847/1538-4365/ac05c0},
  \href {https://ui.adsabs.harvard.edu/abs/2021ApJS..255...30G} {255, 30}

\bibitem[\protect\citeauthoryear{{Graham} et~al.,}{{Graham}
  et~al.}{2020}]{Graham2020}
{Graham} M.~J.,  et~al., 2020, \mn@doi [\prl] {10.1103/PhysRevLett.124.251102},
  \href {https://ui.adsabs.harvard.edu/abs/2020PhRvL.124y1102G} {124, 251102}

\bibitem[\protect\citeauthoryear{{Graham} et~al.,}{{Graham}
  et~al.}{2023}]{Graham2023}
{Graham} M.~J.,  et~al., 2023, \mn@doi [\apj] {10.3847/1538-4357/aca480}, \href
  {https://ui.adsabs.harvard.edu/abs/2023ApJ...942...99G} {942, 99}

\bibitem[\protect\citeauthoryear{{Gravity Collaboration} et~al.,}{{Gravity
  Collaboration} et~al.}{2023}]{Gravity2023}
{Gravity Collaboration} et~al., 2023, \mn@doi [\aap]
  {10.1051/0004-6361/202347416}, \href
  {https://ui.adsabs.harvard.edu/abs/2023A&A...677L..10G} {677, L10}

\bibitem[\protect\citeauthoryear{{Grishin}, {Bobrick}, {Hirai}, {Mandel}  \&
  {Perets}}{{Grishin} et~al.}{2021}]{Grishin2021}
{Grishin} E.,  {Bobrick} A.,  {Hirai} R.,  {Mandel} I.,   {Perets} H.~B.,
  2021, \mn@doi [\mnras] {10.1093/mnras/stab1957}, \href
  {https://ui.adsabs.harvard.edu/abs/2021MNRAS.507..156G} {507, 156}

\bibitem[\protect\citeauthoryear{{Guolo} et~al.,}{{Guolo}
  et~al.}{2024}]{Guolo2024}
{Guolo} M.,  et~al., 2024, \mn@doi [Nature Astronomy]
  {10.1038/s41550-023-02178-4}, \href
  {https://ui.adsabs.harvard.edu/abs/2024NatAs...8..347G} {8, 347}

\bibitem[\protect\citeauthoryear{{Hamann} \& {Ferland}}{{Hamann} \&
  {Ferland}}{1999}]{Hamann1999}
{Hamann} F.,  {Ferland} G.,  1999, \mn@doi [\araa]
  {10.1146/annurev.astro.37.1.487}, \href
  {https://ui.adsabs.harvard.edu/abs/1999ARA&A..37..487H} {37, 487}

\bibitem[\protect\citeauthoryear{{Han}, {Yang}, {Tagawa}, {Jiang}, {Shen},
  {Yun}, {Zhang}  \& {Zhong}}{{Han} et~al.}{2024}]{Han2024}
{Han} W.-B.,  {Yang} S.-C.,  {Tagawa} H.,  {Jiang} Y.,  {Shen} P.,  {Yun} Q.,
  {Zhang} C.,   {Zhong} X.-Y.,  2024, \mn@doi [] {10.48550/arXiv.2401.01743},
  \href {https://ui.adsabs.harvard.edu/abs/2024arXiv240101743H} {p.
  arXiv:2401.01743}

\bibitem[\protect\citeauthoryear{{Helfand}, {White}  \& {Becker}}{{Helfand}
  et~al.}{2015}]{Helfand2015}
{Helfand} D.~J.,  {White} R.~L.,   {Becker} R.~H.,  2015, \mn@doi [\apj]
  {10.1088/0004-637X/801/1/26}, \href
  {https://ui.adsabs.harvard.edu/abs/2015ApJ...801...26H} {801, 26}

\bibitem[\protect\citeauthoryear{{Ho}}{{Ho}}{2008}]{Ho2008}
{Ho} L.~C.,  2008, \mn@doi [\araa] {10.1146/annurev.astro.45.051806.110546},
  \href {https://ui.adsabs.harvard.edu/abs/2008ARA&A..46..475H} {46, 475}

\bibitem[\protect\citeauthoryear{{Hovatta}, {Nieppola}, {Tornikoski},
  {Valtaoja}, {Aller}  \& {Aller}}{{Hovatta} et~al.}{2008}]{Hovatta2008}
{Hovatta} T.,  {Nieppola} E.,  {Tornikoski} M.,  {Valtaoja} E.,  {Aller} M.~F.,
    {Aller} H.~D.,  2008, \mn@doi [\aap] {10.1051/0004-6361:200809806}, \href
  {https://ui.adsabs.harvard.edu/abs/2008A&A...485...51H} {485, 51}

\bibitem[\protect\citeauthoryear{{Hoyle} \& {Lyttleton}}{{Hoyle} \&
  {Lyttleton}}{1939}]{Hoyle1939}
{Hoyle} F.,  {Lyttleton} R.~A.,  1939, \mn@doi [Proceedings of the Cambridge
  Philosophical Society] {10.1017/S0305004100021150}, \href
  {https://ui.adsabs.harvard.edu/abs/1939PCPS...35..405H} {35, 405}

\bibitem[\protect\citeauthoryear{{Huang}, {Lin}  \& {Shields}}{{Huang}
  et~al.}{2023}]{Huang2023}
{Huang} J.,  {Lin} D. N.~C.,   {Shields} G.,  2023, \mn@doi [\mnras]
  {10.1093/mnras/stad2642}, \href
  {https://ui.adsabs.harvard.edu/abs/2023MNRAS.525.5702H} {525, 5702}

\bibitem[\protect\citeauthoryear{{Inoue} \& {Takahara}}{{Inoue} \&
  {Takahara}}{1996}]{Inoue1996}
{Inoue} S.,  {Takahara} F.,  1996, \mn@doi [\apj] {10.1086/177270}, \href
  {https://ui.adsabs.harvard.edu/abs/1996ApJ...463..555I} {463, 555}

\bibitem[\protect\citeauthoryear{{Jermyn}, {Dittmann}, {Cantiello}  \&
  {Perna}}{{Jermyn} et~al.}{2021}]{Jermyn2021}
{Jermyn} A.~S.,  {Dittmann} A.~J.,  {Cantiello} M.,   {Perna} R.,  2021,
  \mn@doi [\apj] {10.3847/1538-4357/abfb67}, \href
  {https://ui.adsabs.harvard.edu/abs/2021ApJ...914..105J} {914, 105}

\bibitem[\protect\citeauthoryear{{Jermyn}, {Dittmann}, {McKernan}, {Ford}  \&
  {Cantiello}}{{Jermyn} et~al.}{2022}]{Jermyn2022}
{Jermyn} A.~S.,  {Dittmann} A.~J.,  {McKernan} B.,  {Ford} K.~E.~S.,
  {Cantiello} M.,  2022, \mn@doi [\apj] {10.3847/1538-4357/ac5d40}, \href
  {https://ui.adsabs.harvard.edu/abs/2022ApJ...929..133J} {929, 133}

\bibitem[\protect\citeauthoryear{{Jiang}, {Stone}  \& {Davis}}{{Jiang}
  et~al.}{2014}]{Jiang2014}
{Jiang} Y.-F.,  {Stone} J.~M.,   {Davis} S.~W.,  2014, \mn@doi [\apj]
  {10.1088/0004-637X/796/2/106}, \href
  {https://ui.adsabs.harvard.edu/abs/2014ApJ...796..106J} {796, 106}

\bibitem[\protect\citeauthoryear{{Kathirgamaraju}, {Li}, {Ryan}  \&
  {Tchekhovskoy}}{{Kathirgamaraju} et~al.}{2023}]{Kathirgamaraju2023}
{Kathirgamaraju} A.,  {Li} H.,  {Ryan} B.~R.,   {Tchekhovskoy} A.,  2023,
  \mn@doi [] {10.48550/arXiv.2311.03571}, \href
  {https://ui.adsabs.harvard.edu/abs/2023arXiv231103571K} {p. arXiv:2311.03571}

\bibitem[\protect\citeauthoryear{{Kato}, {Fukue}  \& {Mineshige}}{{Kato}
  et~al.}{2008}]{Kato2008}
{Kato} S.,  {Fukue} J.,   {Mineshige} S.,  2008, {Black-Hole Accretion Disks
  --- Towards a New Paradigm ---}.
Kyoto, Japan: Kyoto University Press

\bibitem[\protect\citeauthoryear{{Khamitov}, {Bikmaev}, {Gilfanov}, {Sunyaev},
  {Medvedev}  \& {Gorbachev}}{{Khamitov} et~al.}{2023}]{Khamitov2023}
{Khamitov} I.~M.,  {Bikmaev} I.~F.,  {Gilfanov} M.~R.,  {Sunyaev} R.~A.,
  {Medvedev} P.~S.,   {Gorbachev} M.~A.,  2023, \mn@doi []
  {10.48550/arXiv.2309.11308}, \href
  {https://ui.adsabs.harvard.edu/abs/2023arXiv230911308K} {p. arXiv:2309.11308}

\bibitem[\protect\citeauthoryear{{Kimura}, {Murase}  \& {Bartos}}{{Kimura}
  et~al.}{2021}]{Kimura2021}
{Kimura} S.~S.,  {Murase} K.,   {Bartos} I.,  2021, \mn@doi [\apj]
  {10.3847/1538-4357/ac0535}, \href
  {https://ui.adsabs.harvard.edu/abs/2021ApJ...916..111K} {916, 111}

\bibitem[\protect\citeauthoryear{{King}}{{King}}{2003}]{King2003}
{King} A.,  2003, \mn@doi [\apjl] {10.1086/379143}, \href
  {https://ui.adsabs.harvard.edu/abs/2003ApJ...596L..27K} {596, L27}

\bibitem[\protect\citeauthoryear{{Kitaki}, {Mineshige}, {Ohsuga}  \&
  {Kawashima}}{{Kitaki} et~al.}{2021}]{Kitaki2021}
{Kitaki} T.,  {Mineshige} S.,  {Ohsuga} K.,   {Kawashima} T.,  2021, \mn@doi
  [\pasj] {10.1093/pasj/psab011}, \href
  {https://ui.adsabs.harvard.edu/abs/2021PASJ...73..450K} {73, 450}

\bibitem[\protect\citeauthoryear{{Kocsis}, {Yunes}  \& {Loeb}}{{Kocsis}
  et~al.}{2011}]{Kocsis2011}
{Kocsis} B.,  {Yunes} N.,   {Loeb} A.,  2011, \mn@doi [\prd]
  {10.1103/PhysRevD.84.024032}, \href
  {https://ui.adsabs.harvard.edu/abs/2011PhRvD..84b4032K} {84, 024032}

\bibitem[\protect\citeauthoryear{{Lazzati}, {Soares}  \& {Perna}}{{Lazzati}
  et~al.}{2022}]{Lazzati2022}
{Lazzati} D.,  {Soares} G.,   {Perna} R.,  2022, \mn@doi [\apjl]
  {10.3847/2041-8213/ac98ad}, \href
  {https://ui.adsabs.harvard.edu/abs/2022ApJ...938L..18L} {938, L18}

\bibitem[\protect\citeauthoryear{{Lazzati}, {Perna}, {Gompertz}  \&
  {Levan}}{{Lazzati} et~al.}{2023}]{Lazzati2023}
{Lazzati} D.,  {Perna} R.,  {Gompertz} B.~P.,   {Levan} A.~J.,  2023, \mn@doi
  [\apjl] {10.3847/2041-8213/acd18c}, \href
  {https://ui.adsabs.harvard.edu/abs/2023ApJ...950L..20L} {950, L20}

\bibitem[\protect\citeauthoryear{{Lee}, {Cunningham}, {McKee}  \&
  {Klein}}{{Lee} et~al.}{2014}]{Lee2014}
{Lee} A.~T.,  {Cunningham} A.~J.,  {McKee} C.~F.,   {Klein} R.~I.,  2014,
  \mn@doi [\apj] {10.1088/0004-637X/783/1/50}, \href
  {https://ui.adsabs.harvard.edu/abs/2014ApJ...783...50L} {783, 50}

\bibitem[\protect\citeauthoryear{{Levan} et~al.,}{{Levan}
  et~al.}{2023}]{Levan2023}
{Levan} A.~J.,  et~al., 2023, \mn@doi [Nature Astronomy]
  {10.1038/s41550-023-01998-8}, \href
  {https://ui.adsabs.harvard.edu/abs/2023NatAs...7..976L} {7, 976}

\bibitem[\protect\citeauthoryear{{Li} \& {Lai}}{{Li} \&
  {Lai}}{2022}]{LiRixin2022}
{Li} R.,  {Lai} D.,  2022, \mn@doi [\mnras] {10.1093/mnras/stac2577}, \href
  {https://ui.adsabs.harvard.edu/abs/2022MNRAS.517.1602L} {517, 1602}

\bibitem[\protect\citeauthoryear{{Li}, {Dempsey}, {Li}, {Li}  \& {Li}}{{Li}
  et~al.}{2022}]{Li2022}
{Li} Y.-P.,  {Dempsey} A.~M.,  {Li} H.,  {Li} S.,   {Li} J.,  2022, \mn@doi
  [\apjl] {10.3847/2041-8213/ac60fd}, \href
  {https://ui.adsabs.harvard.edu/abs/2022ApJ...928L..19L} {928, L19}

\bibitem[\protect\citeauthoryear{{Li}, {Wang}, {Tang}  \& {Fan}}{{Li}
  et~al.}{2023a}]{LiFan2023}
{Li} Y.-J.,  {Wang} Y.-Z.,  {Tang} S.-P.,   {Fan} Y.-Z.,  2023a, \mn@doi []
  {10.48550/arXiv.2303.02973}, \href
  {https://ui.adsabs.harvard.edu/abs/2023arXiv230302973L} {p. arXiv:2303.02973}

\bibitem[\protect\citeauthoryear{{Li}, {Dempsey}, {Li}, {Lai}  \& {Li}}{{Li}
  et~al.}{2023b}]{Li2023Jiaru}
{Li} J.,  {Dempsey} A.~M.,  {Li} H.,  {Lai} D.,   {Li} S.,  2023b, \mn@doi
  [\apjl] {10.3847/2041-8213/acb934}, \href
  {https://ui.adsabs.harvard.edu/abs/2023ApJ...944L..42L} {944, L42}

\bibitem[\protect\citeauthoryear{{Li}, {Liu}, {Fan}, {Hu}, {Yang}, {Geng}  \&
  {Wu}}{{Li} et~al.}{2023c}]{Li2023}
{Li} F.-L.,  {Liu} Y.,  {Fan} X.,  {Hu} M.-K.,  {Yang} X.,  {Geng} J.-J.,
  {Wu} X.-F.,  2023c, \mn@doi [\apj] {10.3847/1538-4357/acd2d1}, \href
  {https://ui.adsabs.harvard.edu/abs/2023ApJ...950..161L} {950, 161}

\bibitem[\protect\citeauthoryear{{Luo}, {Wu}, {Zhang}, {Wang}, {Ho}  \&
  {Yuan}}{{Luo} et~al.}{2023}]{Luo2023}
{Luo} Y.,  {Wu} X.-J.,  {Zhang} S.-R.,  {Wang} J.-M.,  {Ho} L.~C.,   {Yuan}
  Y.-F.,  2023, \mn@doi [\mnras] {10.1093/mnras/stad2188}, \href
  {https://ui.adsabs.harvard.edu/abs/2023MNRAS.524.6015L} {524, 6015}

\bibitem[\protect\citeauthoryear{{Mauch}, {Murphy}, {Buttery}, {Curran},
  {Hunstead}, {Piestrzynski}, {Robertson}  \& {Sadler}}{{Mauch}
  et~al.}{2003}]{Mauch2003}
{Mauch} T.,  {Murphy} T.,  {Buttery} H.~J.,  {Curran} J.,  {Hunstead} R.~W.,
  {Piestrzynski} B.,  {Robertson} J.~G.,   {Sadler} E.~M.,  2003, \mn@doi
  [\mnras] {10.1046/j.1365-8711.2003.06605.x}, \href
  {https://ui.adsabs.harvard.edu/abs/2003MNRAS.342.1117M} {342, 1117}

\bibitem[\protect\citeauthoryear{{McKernan}, {Ford}, {Lyra}  \&
  {Perets}}{{McKernan} et~al.}{2012}]{McKernan2012}
{McKernan} B.,  {Ford} K.~E.~S.,  {Lyra} W.,   {Perets} H.~B.,  2012, \mn@doi
  [\mnras] {10.1111/j.1365-2966.2012.21486.x}, \href
  {https://ui.adsabs.harvard.edu/abs/2012MNRAS.425..460M} {425, 460}

\bibitem[\protect\citeauthoryear{{McKernan}, {Ford}, {Kocsis}, {Lyra}  \&
  {Winter}}{{McKernan} et~al.}{2014}]{McKernan2014}
{McKernan} B.,  {Ford} K.~E.~S.,  {Kocsis} B.,  {Lyra} W.,   {Winter} L.~M.,
  2014, \mn@doi [\mnras] {10.1093/mnras/stu553}, \href
  {https://ui.adsabs.harvard.edu/abs/2014MNRAS.441..900M} {441, 900}

\bibitem[\protect\citeauthoryear{{McKernan}, {Ford}  \&
  {O'Shaughnessy}}{{McKernan} et~al.}{2020}]{McKernan2020}
{McKernan} B.,  {Ford} K.~E.~S.,   {O'Shaughnessy} R.,  2020, \mn@doi [\mnras]
  {10.1093/mnras/staa2681}, \href
  {https://ui.adsabs.harvard.edu/abs/2020MNRAS.498.4088M} {498, 4088}

\bibitem[\protect\citeauthoryear{{McKernan}, {Ford}, {Cantiello}, {Graham},
  {Jermyn}, {Leigh}, {Ryu}  \& {Stern}}{{McKernan} et~al.}{2022}]{McKernan2022}
{McKernan} B.,  {Ford} K.~E.~S.,  {Cantiello} M.,  {Graham} M.,  {Jermyn}
  A.~S.,  {Leigh} N.~W.~C.,  {Ryu} T.,   {Stern} D.,  2022, \mn@doi [\mnras]
  {10.1093/mnras/stac1310}, \href
  {https://ui.adsabs.harvard.edu/abs/2022MNRAS.514.4102M} {514, 4102}

\bibitem[\protect\citeauthoryear{{Milosavljevi{\'c}}, {Couch}  \&
  {Bromm}}{{Milosavljevi{\'c}} et~al.}{2009}]{Milosavljevic2009}
{Milosavljevi{\'c}} M.,  {Couch} S.~M.,   {Bromm} V.,  2009, \mn@doi [\apjl]
  {10.1088/0004-637X/696/2/L146}, \href
  {https://ui.adsabs.harvard.edu/abs/2009ApJ...696L.146M} {696, L146}

\bibitem[\protect\citeauthoryear{{Mooley} et~al.,}{{Mooley}
  et~al.}{2016}]{Mooley2016}
{Mooley} K.~P.,  et~al., 2016, \mn@doi [\apj] {10.3847/0004-637X/818/2/105},
  \href {https://ui.adsabs.harvard.edu/abs/2016ApJ...818..105M} {818, 105}

\bibitem[\protect\citeauthoryear{{Morton}, {Rinaldi}, {Torres-Orjuela},
  {Derdzinski}, {Vaccaro}  \& {Del Pozzo}}{{Morton} et~al.}{2023}]{Morton2023}
{Morton} S.~L.,  {Rinaldi} S.,  {Torres-Orjuela} A.,  {Derdzinski} A.,
  {Vaccaro} M.~P.,   {Del Pozzo} W.,  2023, \mn@doi [\prd]
  {10.1103/PhysRevD.108.123039}, \href
  {https://ui.adsabs.harvard.edu/abs/2023PhRvD.108l3039M} {108, 123039}

\bibitem[\protect\citeauthoryear{{Muchotrzeb} \& {Paczynski}}{{Muchotrzeb} \&
  {Paczynski}}{1982}]{Muchotrzeb1982}
{Muchotrzeb} B.,  {Paczynski} B.,  1982, \actaa, \href
  {https://ui.adsabs.harvard.edu/abs/1982AcA....32....1M} {32, 1}

\bibitem[\protect\citeauthoryear{{Murphy} et~al.,}{{Murphy}
  et~al.}{2021}]{Murphy2021}
{Murphy} T.,  et~al., 2021, \mn@doi [\pasa] {10.1017/pasa.2021.44}, \href
  {https://ui.adsabs.harvard.edu/abs/2021PASA...38...54M} {38, e054}

\bibitem[\protect\citeauthoryear{{Narayan} \& {Yi}}{{Narayan} \&
  {Yi}}{1995}]{Narayan1995ADAF}
{Narayan} R.,  {Yi} I.,  1995, \mn@doi [\apj] {10.1086/176343}, \href
  {https://ui.adsabs.harvard.edu/abs/1995ApJ...452..710N} {452, 710}

\bibitem[\protect\citeauthoryear{{Narayan}, {Yi}  \& {Mahadevan}}{{Narayan}
  et~al.}{1995}]{Narayan1995}
{Narayan} R.,  {Yi} I.,   {Mahadevan} R.,  1995, \mn@doi [\nat]
  {10.1038/374623a0}, \href
  {https://ui.adsabs.harvard.edu/abs/1995Natur.374..623N} {374, 623}

\bibitem[\protect\citeauthoryear{{Nyland} et~al.,}{{Nyland}
  et~al.}{2020}]{Nyland2020}
{Nyland} K.,  et~al., 2020, \mn@doi [\apj] {10.3847/1538-4357/abc341}, \href
  {https://ui.adsabs.harvard.edu/abs/2020ApJ...905...74N} {905, 74}

\bibitem[\protect\citeauthoryear{{Ofek} et~al.,}{{Ofek}
  et~al.}{2021}]{Ofek2021}
{Ofek} E.~O.,  et~al., 2021, \mn@doi [\apj] {10.3847/1538-4357/ac24fc}, \href
  {https://ui.adsabs.harvard.edu/abs/2021ApJ...922..247O} {922, 247}

\bibitem[\protect\citeauthoryear{{Ohsuga}, {Mori}, {Nakamoto}  \&
  {Mineshige}}{{Ohsuga} et~al.}{2005}]{Ohsuga2005}
{Ohsuga} K.,  {Mori} M.,  {Nakamoto} T.,   {Mineshige} S.,  2005, \mn@doi
  [\apj] {10.1086/430728}, \href
  {https://ui.adsabs.harvard.edu/abs/2005ApJ...628..368O} {628, 368}

\bibitem[\protect\citeauthoryear{{Paliya}, {Dom{\'\i}nguez}, {Ajello},
  {Franckowiak}  \& {Hartmann}}{{Paliya} et~al.}{2019}]{Paliya2019}
{Paliya} V.~S.,  {Dom{\'\i}nguez} A.,  {Ajello} M.,  {Franckowiak} A.,
  {Hartmann} D.,  2019, \mn@doi [\apjl] {10.3847/2041-8213/ab398a}, \href
  {https://ui.adsabs.harvard.edu/abs/2019ApJ...882L...3P} {882, L3}

\bibitem[\protect\citeauthoryear{{Palmese}, {Fishbach}, {Burke}, {Annis}  \&
  {Liu}}{{Palmese} et~al.}{2021}]{Palmese2021}
{Palmese} A.,  {Fishbach} M.,  {Burke} C.~J.,  {Annis} J.,   {Liu} X.,  2021,
  \mn@doi [\apjl] {10.3847/2041-8213/ac0883}, \href
  {https://ui.adsabs.harvard.edu/abs/2021ApJ...914L..34P} {914, L34}

\bibitem[\protect\citeauthoryear{{Pan} \& {Yang}}{{Pan} \&
  {Yang}}{2021}]{Pan2021}
{Pan} Z.,  {Yang} H.,  2021, \mn@doi [\apj] {10.3847/1538-4357/ac249c}, \href
  {https://ui.adsabs.harvard.edu/abs/2021ApJ...923..173P} {923, 173}

\bibitem[\protect\citeauthoryear{{Park} \& {Trippe}}{{Park} \&
  {Trippe}}{2017}]{Park2017}
{Park} J.,  {Trippe} S.,  2017, \mn@doi [\apj] {10.3847/1538-4357/834/2/157},
  \href {https://ui.adsabs.harvard.edu/abs/2017ApJ...834..157P} {834, 157}

\bibitem[\protect\citeauthoryear{{Perna}, {Lazzati}  \& {Cantiello}}{{Perna}
  et~al.}{2021a}]{Perna2021EMC}
{Perna} R.,  {Lazzati} D.,   {Cantiello} M.,  2021a, \mn@doi [\apjl]
  {10.3847/2041-8213/abd319}, \href
  {https://ui.adsabs.harvard.edu/abs/2021ApJ...906L...7P} {906, L7}

\bibitem[\protect\citeauthoryear{{Perna}, {Tagawa}, {Haiman}  \&
  {Bartos}}{{Perna} et~al.}{2021b}]{Perna2021AIC}
{Perna} R.,  {Tagawa} H.,  {Haiman} Z.,   {Bartos} I.,  2021b, \mn@doi [\apj]
  {10.3847/1538-4357/abfdb4}, \href
  {https://ui.adsabs.harvard.edu/abs/2021ApJ...915...10P} {915, 10}

\bibitem[\protect\citeauthoryear{{Qian}, {Li}  \& {Lai}}{{Qian}
  et~al.}{2024}]{Qian2024}
{Qian} K.,  {Li} J.,   {Lai} D.,  2024, \mn@doi [\apj]
  {10.3847/1538-4357/ad1b53}, \href
  {https://ui.adsabs.harvard.edu/abs/2024ApJ...962..143Q} {962, 143}

\bibitem[\protect\citeauthoryear{{Ray}, {Lazzati}  \& {Perna}}{{Ray}
  et~al.}{2023}]{Ray2023}
{Ray} M.,  {Lazzati} D.,   {Perna} R.,  2023, \mn@doi [\mnras]
  {10.1093/mnras/stad816}, \href
  {https://ui.adsabs.harvard.edu/abs/2023MNRAS.521.4233R} {521, 4233}

\bibitem[\protect\citeauthoryear{{Rees}, {Begelman}, {Blandford}  \&
  {Phinney}}{{Rees} et~al.}{1982}]{Rees1982}
{Rees} M.~J.,  {Begelman} M.~C.,  {Blandford} R.~D.,   {Phinney} E.~S.,  1982,
  \mn@doi [\nat] {10.1038/295017a0}, \href
  {https://ui.adsabs.harvard.edu/abs/1982Natur.295...17R} {295, 17}

\bibitem[\protect\citeauthoryear{{Ren}, {Chen}, {Wang}  \& {Dai}}{{Ren}
  et~al.}{2022}]{Ren2022}
{Ren} J.,  {Chen} K.,  {Wang} Y.,   {Dai} Z.-G.,  2022, \mn@doi [\apjl]
  {10.3847/2041-8213/aca025}, \href
  {https://ui.adsabs.harvard.edu/abs/2022ApJ...940L..44R} {940, L44}

\bibitem[\protect\citeauthoryear{{Rom}, {Sari}  \& {Lai}}{{Rom}
  et~al.}{2024}]{Rom2024}
{Rom} B.,  {Sari} R.,   {Lai} D.,  2024, \mn@doi [\apj]
  {10.3847/1538-4357/ad284b}, \href
  {https://ui.adsabs.harvard.edu/abs/2024ApJ...964...43R} {964, 43}

\bibitem[\protect\citeauthoryear{{Rowan}, {Boekholt}, {Kocsis}  \&
  {Haiman}}{{Rowan} et~al.}{2023}]{Rowan2023}
{Rowan} C.,  {Boekholt} T.,  {Kocsis} B.,   {Haiman} Z.,  2023, \mn@doi
  [\mnras] {10.1093/mnras/stad1926}, \href
  {https://ui.adsabs.harvard.edu/abs/2023MNRAS.524.2770R} {524, 2770}

\bibitem[\protect\citeauthoryear{{Samsing} et~al.,}{{Samsing}
  et~al.}{2022}]{Samsing2022}
{Samsing} J.,  et~al., 2022, \mn@doi [\nat] {10.1038/s41586-021-04333-1}, \href
  {https://ui.adsabs.harvard.edu/abs/2022Natur.603..237S} {603, 237}

\bibitem[\protect\citeauthoryear{{Shakura} \& {Sunyaev}}{{Shakura} \&
  {Sunyaev}}{1973}]{Shakura1973}
{Shakura} N.~I.,  {Sunyaev} R.~A.,  1973, \aap, \href
  {https://ui.adsabs.harvard.edu/abs/1973A&A....24..337S} {24, 337}

\bibitem[\protect\citeauthoryear{{Shimwell} et~al.,}{{Shimwell}
  et~al.}{2022}]{Shimwell2022}
{Shimwell} T.~W.,  et~al., 2022, \mn@doi [\aap] {10.1051/0004-6361/202142484},
  \href {https://ui.adsabs.harvard.edu/abs/2022A&A...659A...1S} {659, A1}

\bibitem[\protect\citeauthoryear{{Sirko} \& {Goodman}}{{Sirko} \&
  {Goodman}}{2003}]{Sirko2003}
{Sirko} E.,  {Goodman} J.,  2003, \mn@doi [\mnras]
  {10.1046/j.1365-8711.2003.06431.x}, \href
  {https://ui.adsabs.harvard.edu/abs/2003MNRAS.341..501S} {341, 501}

\bibitem[\protect\citeauthoryear{{Stone}, {Metzger}  \& {Haiman}}{{Stone}
  et~al.}{2017}]{Stone2017}
{Stone} N.~C.,  {Metzger} B.~D.,   {Haiman} Z.,  2017, \mn@doi [\mnras]
  {10.1093/mnras/stw2260}, \href
  {https://ui.adsabs.harvard.edu/abs/2017MNRAS.464..946S} {464, 946}

\bibitem[\protect\citeauthoryear{{Svensson} \& {Zdziarski}}{{Svensson} \&
  {Zdziarski}}{1994}]{Svensson1994}
{Svensson} R.,  {Zdziarski} A.~A.,  1994, \mn@doi [\apj] {10.1086/174934},
  \href {https://ui.adsabs.harvard.edu/abs/1994ApJ...436..599S} {436, 599}

\bibitem[\protect\citeauthoryear{{Tagawa}, {Haiman}  \& {Kocsis}}{{Tagawa}
  et~al.}{2020}]{Tagawa2020}
{Tagawa} H.,  {Haiman} Z.,   {Kocsis} B.,  2020, \mn@doi [\apj]
  {10.3847/1538-4357/ab9b8c}, \href
  {https://ui.adsabs.harvard.edu/abs/2020ApJ...898...25T} {898, 25}

\bibitem[\protect\citeauthoryear{{Tagawa}, {Kimura}, {Haiman}, {Perna},
  {Tanaka}  \& {Bartos}}{{Tagawa} et~al.}{2022}]{Tagawa2022}
{Tagawa} H.,  {Kimura} S.~S.,  {Haiman} Z.,  {Perna} R.,  {Tanaka} H.,
  {Bartos} I.,  2022, \mn@doi [\apj] {10.3847/1538-4357/ac45f8}, \href
  {https://ui.adsabs.harvard.edu/abs/2022ApJ...927...41T} {927, 41}

\bibitem[\protect\citeauthoryear{{Tagawa}, {Kimura}, {Haiman}, {Perna}  \&
  {Bartos}}{{Tagawa} et~al.}{2023a}]{Tagawa2023Shock}
{Tagawa} H.,  {Kimura} S.~S.,  {Haiman} Z.,  {Perna} R.,   {Bartos} I.,  2023a,
  \mn@doi [] {10.48550/arXiv.2310.18392}, \href
  {https://ui.adsabs.harvard.edu/abs/2023arXiv231018392T} {p. arXiv:2310.18392}

\bibitem[\protect\citeauthoryear{{Tagawa}, {Kimura}, {Haiman}, {Perna}  \&
  {Bartos}}{{Tagawa} et~al.}{2023b}]{Tagawa2023BH}
{Tagawa} H.,  {Kimura} S.~S.,  {Haiman} Z.,  {Perna} R.,   {Bartos} I.,  2023b,
  \mn@doi [\apjl] {10.3847/2041-8213/acc103}, \href
  {https://ui.adsabs.harvard.edu/abs/2023ApJ...946L...3T} {946, L3}

\bibitem[\protect\citeauthoryear{{Tagawa}, {Kimura}, {Haiman}, {Perna}  \&
  {Bartos}}{{Tagawa} et~al.}{2023c}]{Tagawa2023BBH}
{Tagawa} H.,  {Kimura} S.~S.,  {Haiman} Z.,  {Perna} R.,   {Bartos} I.,  2023c,
  \mn@doi [\apj] {10.3847/1538-4357/acc4bb}, \href
  {https://ui.adsabs.harvard.edu/abs/2023ApJ...950...13T} {950, 13}

\bibitem[\protect\citeauthoryear{{Takeuchi}, {Mineshige}  \&
  {Ohsuga}}{{Takeuchi} et~al.}{2009}]{Takeuchi2009}
{Takeuchi} S.,  {Mineshige} S.,   {Ohsuga} K.,  2009, \mn@doi [\pasj]
  {10.1093/pasj/61.4.783}, \href
  {https://ui.adsabs.harvard.edu/abs/2009PASJ...61..783T} {61, 783}

\bibitem[\protect\citeauthoryear{{Tanaka}, {Takeuchi}  \& {Ward}}{{Tanaka}
  et~al.}{2002}]{Tanaka2002}
{Tanaka} H.,  {Takeuchi} T.,   {Ward} W.~R.,  2002, \mn@doi [\apj]
  {10.1086/324713}, \href
  {https://ui.adsabs.harvard.edu/abs/2002ApJ...565.1257T} {565, 1257}

\bibitem[\protect\citeauthoryear{{Toomre}}{{Toomre}}{1964}]{Toomre1964}
{Toomre} A.,  1964, \mn@doi [\apj] {10.1086/147861}, \href
  {https://ui.adsabs.harvard.edu/abs/1964ApJ...139.1217T} {139, 1217}

\bibitem[\protect\citeauthoryear{{Uchiyama}, {Aharonian}, {Tanaka}, {Takahashi}
   \& {Maeda}}{{Uchiyama} et~al.}{2007}]{Uchiyama2007}
{Uchiyama} Y.,  {Aharonian} F.~A.,  {Tanaka} T.,  {Takahashi} T.,   {Maeda} Y.,
   2007, \mn@doi [\nat] {10.1038/nature06210}, \href
  {https://ui.adsabs.harvard.edu/abs/2007Natur.449..576U} {449, 576}

\bibitem[\protect\citeauthoryear{{Wandel}}{{Wandel}}{1984}]{Wandel1984}
{Wandel} A.,  1984, \mn@doi [\mnras] {10.1093/mnras/207.4.861}, \href
  {https://ui.adsabs.harvard.edu/abs/1984MNRAS.207..861W} {207, 861}

\bibitem[\protect\citeauthoryear{{Wang} \& {Zhou}}{{Wang} \&
  {Zhou}}{1999}]{Wang1999}
{Wang} J.-M.,  {Zhou} Y.-Y.,  1999, \mn@doi [\apj] {10.1086/307080}, \href
  {https://ui.adsabs.harvard.edu/abs/1999ApJ...516..420W} {516, 420}

\bibitem[\protect\citeauthoryear{{Wang}, {Chen}  \& {Hu}}{{Wang}
  et~al.}{2006}]{Wang2006}
{Wang} J.-M.,  {Chen} Y.-M.,   {Hu} C.,  2006, \mn@doi [\apjl]
  {10.1086/500557}, \href
  {https://ui.adsabs.harvard.edu/abs/2006ApJ...637L..85W} {637, L85}

\bibitem[\protect\citeauthoryear{{Wang}, {Yan}, {Gao}, {Hu}, {Li}  \&
  {Zhang}}{{Wang} et~al.}{2010}]{Wang2010}
{Wang} J.-M.,  {Yan} C.-S.,  {Gao} H.-Q.,  {Hu} C.,  {Li} Y.-R.,   {Zhang} S.,
  2010, \mn@doi [\apjl] {10.1088/2041-8205/719/2/L148}, \href
  {https://ui.adsabs.harvard.edu/abs/2010ApJ...719L.148W} {719, L148}

\bibitem[\protect\citeauthoryear{{Wang} et~al.,}{{Wang}
  et~al.}{2011}]{Wang2011}
{Wang} J.-M.,  et~al., 2011, \mn@doi [\apj] {10.1088/0004-637X/739/1/3}, \href
  {https://ui.adsabs.harvard.edu/abs/2011ApJ...739....3W} {739, 3}

\bibitem[\protect\citeauthoryear{{Wang}, {Du}, {Baldwin}, {Ge}, {Hu}  \&
  {Ferland}}{{Wang} et~al.}{2012}]{Wang2012}
{Wang} J.-M.,  {Du} P.,  {Baldwin} J.~A.,  {Ge} J.-Q.,  {Hu} C.,   {Ferland}
  G.~J.,  2012, \mn@doi [\apj] {10.1088/0004-637X/746/2/137}, \href
  {https://ui.adsabs.harvard.edu/abs/2012ApJ...746..137W} {746, 137}

\bibitem[\protect\citeauthoryear{{Wang}, {Liu}, {Ho}  \& {Du}}{{Wang}
  et~al.}{2021a}]{PaperI}
{Wang} J.-M.,  {Liu} J.-R.,  {Ho} L.~C.,   {Du} P.,  2021a, \mn@doi [\apjl]
  {10.3847/2041-8213/abee81}, \href
  {https://ui.adsabs.harvard.edu/abs/2021ApJ...911L..14W} {911, L14}

\bibitem[\protect\citeauthoryear{{Wang}, {Liu}, {Ho}, {Li}  \& {Du}}{{Wang}
  et~al.}{2021b}]{PaperII}
{Wang} J.-M.,  {Liu} J.-R.,  {Ho} L.~C.,  {Li} Y.-R.,   {Du} P.,  2021b,
  \mn@doi [\apjl] {10.3847/2041-8213/ac0b46}, \href
  {https://ui.adsabs.harvard.edu/abs/2021ApJ...916L..17W} {916, L17}

\bibitem[\protect\citeauthoryear{{Wang}, {Lazzati}  \& {Perna}}{{Wang}
  et~al.}{2022}]{Wang2022}
{Wang} Y.-H.,  {Lazzati} D.,   {Perna} R.,  2022, \mn@doi [\mnras]
  {10.1093/mnras/stac1968}, \href
  {https://ui.adsabs.harvard.edu/abs/2022MNRAS.516.5935W} {516, 5935}

\bibitem[\protect\citeauthoryear{{Wang}, {Ma}  \& {Wu}}{{Wang}
  et~al.}{2023a}]{WangMengye2023}
{Wang} M.,  {Ma} Y.,   {Wu} Q.,  2023a, \mn@doi [\mnras]
  {10.1093/mnras/stad422}, \href
  {https://ui.adsabs.harvard.edu/abs/2023MNRAS.520.4502W} {520, 4502}

\bibitem[\protect\citeauthoryear{{Wang} et~al.,}{{Wang}
  et~al.}{2023b}]{Wang2023}
{Wang} J.-M.,  et~al., 2023b, \mn@doi [\apj] {10.3847/1538-4357/acdf48}, \href
  {https://ui.adsabs.harvard.edu/abs/2023ApJ...954...84W} {954, 84}

\bibitem[\protect\citeauthoryear{{Wang}, {Liu}, {Li}, {Songsheng}, {Yuan}  \&
  {Ho}}{{Wang} et~al.}{2023c}]{PaperIII}
{Wang} J.-M.,  {Liu} J.-R.,  {Li} Y.-R.,  {Songsheng} Y.-Y.,  {Yuan} Y.-F.,
  {Ho} L.~C.,  2023c, \mn@doi [\apjl] {10.3847/2041-8213/ad0bd9}, \href
  {https://ui.adsabs.harvard.edu/abs/2023ApJ...958L..40W} {958, L40}

\bibitem[\protect\citeauthoryear{{Wang}, {Zhu}  \& {Lin}}{{Wang}
  et~al.}{2024}]{Wang2024}
{Wang} Y.,  {Zhu} Z.,   {Lin} D. N.~C.,  2024, \mn@doi [\mnras]
  {10.1093/mnras/stae321}, \href
  {https://ui.adsabs.harvard.edu/abs/2024MNRAS.528.4958W} {528, 4958}

\bibitem[\protect\citeauthoryear{{Witzel} et~al.,}{{Witzel}
  et~al.}{2021}]{Witzel2021}
{Witzel} G.,  et~al., 2021, \mn@doi [\apj] {10.3847/1538-4357/ac0891}, \href
  {https://ui.adsabs.harvard.edu/abs/2021ApJ...917...73W} {917, 73}

\bibitem[\protect\citeauthoryear{{Yang}, {Yuan}, {Ohsuga}  \& {Bu}}{{Yang}
  et~al.}{2014}]{Yang2014}
{Yang} X.-H.,  {Yuan} F.,  {Ohsuga} K.,   {Bu} D.-F.,  2014, \mn@doi [\apj]
  {10.1088/0004-637X/780/1/79}, \href
  {https://ui.adsabs.harvard.edu/abs/2014ApJ...780...79Y} {780, 79}

\bibitem[\protect\citeauthoryear{{Yang} et~al.,}{{Yang}
  et~al.}{2019}]{Yang2019}
{Yang} Y.,  et~al., 2019, \mn@doi [\prl] {10.1103/PhysRevLett.123.181101},
  \href {https://ui.adsabs.harvard.edu/abs/2019PhRvL.123r1101Y} {123, 181101}

\bibitem[\protect\citeauthoryear{{Yang}, {Bartos}, {Fragione}, {Haiman},
  {Kowalski}, {M{\'a}rka}, {Perna}  \& {Tagawa}}{{Yang}
  et~al.}{2022}]{Yang2022}
{Yang} Y.,  {Bartos} I.,  {Fragione} G.,  {Haiman} Z.,  {Kowalski} M.,
  {M{\'a}rka} S.,  {Perna} R.,   {Tagawa} H.,  2022, \mn@doi [\apjl]
  {10.3847/2041-8213/ac7c0b}, \href
  {https://ui.adsabs.harvard.edu/abs/2022ApJ...933L..28Y} {933, L28}

\bibitem[\protect\citeauthoryear{{Yi} \& {Cheng}}{{Yi} \&
  {Cheng}}{2019}]{Yi2019}
{Yi} S.-X.,  {Cheng} K.~S.,  2019, \mn@doi [\apjl] {10.3847/2041-8213/ab459a},
  \href {https://ui.adsabs.harvard.edu/abs/2019ApJ...884L..12Y} {884, L12}

\bibitem[\protect\citeauthoryear{{Yuan}, {Murase}, {Guetta}, {Pe'er}, {Bartos}
  \& {M{\'e}sz{\'a}ros}}{{Yuan} et~al.}{2022}]{Yuan2022}
{Yuan} C.,  {Murase} K.,  {Guetta} D.,  {Pe'er} A.,  {Bartos} I.,
  {M{\'e}sz{\'a}ros} P.,  2022, \mn@doi [\apj] {10.3847/1538-4357/ac6ddf},
  \href {https://ui.adsabs.harvard.edu/abs/2022ApJ...932...80Y} {932, 80}

\bibitem[\protect\citeauthoryear{{Zhang}, {Shu}, {Sun}, {Yang}, {Jiang}, {Dou},
  {Wang}  \& {Wang}}{{Zhang} et~al.}{2022}]{Zhang2022}
{Zhang} F.,  {Shu} X.,  {Sun} L.,  {Yang} L.,  {Jiang} N.,  {Dou} L.,  {Wang}
  J.,   {Wang} T.,  2022, \mn@doi [\apj] {10.3847/1538-4357/ac8a9a}, \href
  {https://ui.adsabs.harvard.edu/abs/2022ApJ...938...43Z} {938, 43}

\bibitem[\protect\citeauthoryear{{Zhang}, {Luo}, {Wu}, {Wang}, {Ho}  \&
  {Yuan}}{{Zhang} et~al.}{2023}]{Zhang2023}
{Zhang} S.-R.,  {Luo} Y.,  {Wu} X.-J.,  {Wang} J.-M.,  {Ho} L.~C.,   {Yuan}
  Y.-F.,  2023, \mn@doi [\mnras] {10.1093/mnras/stad1855}, \href
  {https://ui.adsabs.harvard.edu/abs/2023MNRAS.524..940Z} {524, 940}

\bibitem[\protect\citeauthoryear{{Zhu}, {Zhang}, {Yu}  \& {Gao}}{{Zhu}
  et~al.}{2021a}]{Zhu2021NSmerger}
{Zhu} J.-P.,  {Zhang} B.,  {Yu} Y.-W.,   {Gao} H.,  2021a, \mn@doi [\apjl]
  {10.3847/2041-8213/abd412}, \href
  {https://ui.adsabs.harvard.edu/abs/2021ApJ...906L..11Z} {906, L11}

\bibitem[\protect\citeauthoryear{{Zhu}, {Yang}, {Zhang}, {Liu}, {Yu}  \&
  {Gao}}{{Zhu} et~al.}{2021b}]{Zhu2021WD}
{Zhu} J.-P.,  {Yang} Y.-P.,  {Zhang} B.,  {Liu} L.-D.,  {Yu} Y.-W.,   {Gao} H.,
   2021b, \mn@doi [\apjl] {10.3847/2041-8213/abff5a}, \href
  {https://ui.adsabs.harvard.edu/abs/2021ApJ...914L..19Z} {914, L19}

\bibitem[\protect\citeauthoryear{{de Menezes}, {Nemmen}, {Finke}, {Almeida}  \&
  {Rani}}{{de Menezes} et~al.}{2020}]{Menezes2020}
{de Menezes} R.,  {Nemmen} R.,  {Finke} J.~D.,  {Almeida} I.,   {Rani} B.,
  2020, \mn@doi [\mnras] {10.1093/mnras/staa083}, \href
  {https://ui.adsabs.harvard.edu/abs/2020MNRAS.492.4120D} {492, 4120}

\makeatother
\end{thebibliography}
\bibliographystyle{mnras}
\end{document}